\pdfoutput=1
\documentclass[11pt]{article}

\usepackage[margin=1in]{geometry}
\usepackage{setspace}
\usepackage{amsmath,amssymb,amsfonts,amsthm}
\usepackage{mathtools}
\usepackage{bm}
\usepackage{enumitem}
\usepackage{booktabs}
\usepackage{array}
\usepackage[font=small,labelfont=bf,labelsep=period,skip=6pt]{caption}
\usepackage{siunitx}
\usepackage{graphicx}
\usepackage{xcolor}
\usepackage{comment}
\usepackage{algorithm,algpseudocode}
\usepackage{natbib}
\setcitestyle{round,semicolon}
\usepackage{float}
\usepackage{placeins}
\usepackage{xurl}
\usepackage{hyperref}
\hypersetup{
  pdftitle={Designing Customer Feedback for Subscription Boxes},
  pdfauthor={Renjun Hu, Hyun-Soo Ahn, Stefanus Jasin},
  pdfkeywords={subscription boxes, customer feedback, Bayesian learning, value of information, top-K selection}
}
\hypersetup{
  colorlinks=true,
  linkcolor=blue!60!black,
  citecolor=blue!60!black,
  urlcolor=blue!60!black
}
\newtheorem{theorem}{Theorem}
\newtheorem{proposition}[theorem]{Proposition}
\newtheorem{lemma}[theorem]{Lemma}
\newtheorem{corollary}[theorem]{Corollary}
\theoremstyle{definition}
\newtheorem{definition}{Definition}

\theoremstyle{remark}

\newcommand{\R}{\mathbb{R}}

\newcommand{\E}{\mathbb{E}}

\newcommand{\1}{\mathbf{1}}
\newcommand{\w}{\mathbf{w}}

\newcommand{\trace}{\mathrm{tr}}
\newcommand{\diag}{\mathrm{diag}}

\newcommand{\bmu}{\boldsymbol{\mu}}

\newcommand{\bSigma}{\boldsymbol{\Sigma}}
\newcommand{\N}{\mathbf{N}}

\newcommand{\cD}{\mathcal{D}}

\title{\vspace{-1em}\textbf{Designing Customer Feedback for Subscription Boxes}}
\author{
Renjun Hu \qquad Hyun-Soo Ahn \qquad Stefanus Jasin \\[0.6em]
Ross School of Business, University of Michigan \\[0.25em]
\small \texttt{renjunhu@umich.edu} \qquad \texttt{hsahn@umich.edu} \qquad \texttt{sjasin@umich.edu}
}
\date{October 2026}

\begin{document}
\maketitle

\begin{abstract}
\noindent\textit{Problem definition}: Subscription box services often
ask customers for ratings before choosing what to send. Responses are
scarce, and longer questionnaires and finer scales lower completion. We
study which candidate items a questionnaire should ask about, and in what
format, when the platform then selects a fixed number of them. A design
is judged by the expected value lost by not selecting the truly best
items.

\noindent\textit{Methodology/results}: We approximate each ordinal rating
by a Gaussian signal carrying its information at the cutoff between
selected and unselected items. As equal numbers of responses per item grow, the ratio of this proxy's loss to the exact loss tends to one. With 1 to 100 responses per item, the two are within about 7 percent. Averaging the information over the prior instead can misstate the loss by about half. For an item far from the cutoff,
the value of responses grows slowly at first and then faster, so our
envelope rule values responses in blocks. In simulations with a correct
prior it lowers the loss by 15 to 81 percent relative to an even spread,
and on MovieLens ratings by 6 to 25 percent. Under the proxy, with responses allocated optimally, formats are ranked exactly by one number, the
expected information per customer asked. Choosing the format this way
comes within about 2 percent of the best on average.

\noindent\textit{Managerial implications}: Feedback should go to items
whose place in the selection is unclear. Planning helps most when the
platform's prior beliefs are informative and right. A platform unsure of
its prior should widen it and send the questionnaire in waves.
Calibrated on MovieLens ratings, a five-point answer carries about 1.7
times the information of a like/dislike answer, so at equal length the
finer scale is better when it cuts completion by less than about 40
percent.
\end{abstract}

\section{Introduction}
\label{sec:intro}

A subscription box platform decides what goes into the boxes it sends,
such as the recipes on a meal kit's menu or the products in a beauty
box. Such platforms sell curation subscriptions, which differ from
replenishment subscriptions that refill products the customer already
knows \citep{ChenEtAl2018}.
The global subscription-box market was valued at about \$30 billion in
2024 \citep{ResearchAndMarkets2025}. Customers pay for a convenient
selection and the chance to discover items they like. A box that
feels random, repetitive, or poorly matched gives them less reason to
stay subscribed \citep{ChenEtAl2018}. In about 52{,}000 orders at a
curation service, a one-standard-deviation rise in the share of items
repeated from a customer's previous four orders was associated with 12 to 37
percent lower odds of ordering again, depending on the customer segment
(Appendix~\ref{app:retention-drivers}).
The selection is hard to get right because it is often made for
customers the platform knows little about, such as those on a newly
launched plan. We focus on platforms that make one selection for the
group and so need to learn which items the group likes on average.
Platforms already ask customers what they like
\citep{StitchFixBestFix2025,HungryrootPersonalized2025,HungryrootPreferences2025}.

A platform must choose $K$ of $L$ candidate items, such as the contents
of a box, for a target customer population. Customers' utilities from an
item differ, but the platform cannot tell customers apart before the
selection. We therefore call the average utility of item $i$ across the
population its value $w_i$. If the
platform knew these values, it would choose the $K$ items with the
highest values. Lacking them, it can ask customers for feedback before this
top-$K$ decision, for example through a short questionnaire
emailed after a delivery. Such feedback is scarce. Each customer can be shown only a
few questions, and most do not answer at all. Survey response rates
average about \(29\) percent online and \(13\) percent in apps
\citep{Malnik2025}, and longer questionnaires lower completion further
\citep{Galesic2009}. In a Netflix test, a thumbs-up/down prompt drew
about three times as many ratings as a five-star one
\citep{Netflix2017}.

The platform therefore faces two design questions, which we treat separately.
\begin{enumerate}[leftmargin=*]
\item \textbf{Where should feedback be collected?} Given a limited
number of responses, which items should be asked about?
\item \textbf{What feedback format should be used?} How many questions
should each customer receive, and how fine should each rating scale
be?
\end{enumerate}
The format determines the number of responses and the information in
each, and the allocation assigns them to items. Both
are planned once, before the questionnaire is sent. Learning from
the responses is standard. Each
response is an ordinal rating, such as thumbs up or down or one to five
stars. The platform updates its prior beliefs by Bayes' rule and
selects the $K$ items with the highest estimated values.

The objective is the quality of the eventual top-$K$ decision, not
prediction accuracy or the number of responses collected. Feedback is
useful to the extent that it improves which items are selected. Consider a platform choosing five recipes out of forty.
More responses about the most and least popular recipes would sharpen
their estimates without changing which five are chosen. Responses
about the few recipes near fifth place can change that choice. We therefore evaluate a feedback design by its
\emph{decision loss}, the value the platform gives up because it selects
items without knowing their true values.

\paragraph{Main results and contributions.}
For tractability, we replace each rating by a
Gaussian signal with the same information and check this proxy against
the ordinal model. We give one result per design
question and one on the proxy, and test them in
simulation and, for allocation, on real ratings.
\begin{enumerate}[leftmargin=*]
\item \textbf{Feedback is worth most for items whose place in the
selection is unclear.}
We call the boundary between the $K$ items with the highest values and
the rest the cutoff. A response matters more when its item is more
likely to be near the cutoff and when the platform is still uncertain
about it. Asking about the most uncertain items ignores the
cutoff, and in our experiments it does no better than spreading
responses evenly. For an item far from the cutoff, the value of
responses grows slowly over the first few and faster after that, so a plan
that adds responses one at a time may never ask about it. We propose an
envelope rule that values each item's responses in blocks.

With the cutoffs held fixed, the rule falls short of the best plan for those cutoffs by at most what one item's responses can add. With a large budget, the best plan gives each item a share of responses
proportional to the square root of how likely it is to be near the
cutoff. Planning gains most when few items are likely to be near it,
as under an informative prior. In simulations with a correct prior, the envelope rule lowers
the decision loss by 15 to 81 percent relative to an even spread. Most
of this gain comes from favoring items near the cutoff, as the
square-root rule does. A platform unsure of its prior should widen it
and send the questionnaire in waves.

\item \textbf{Formats can be compared through a single measure of
information.}
Longer questionnaires and finer scales provide more information per
respondent but lower the completion rate. Because the proxy
gives every item the same information per rating, a change of scale
multiplies the information about all items by a common factor. Under
the proxy, formats are then ranked exactly by one number, the expected
information per customer asked, when responses are allocated
optimally. The envelope rule's loss also depends on the format only
through this number. Calibrated on MovieLens ratings, at equal length a five-point scale is
better when it cuts completion by less than about 40 percent. If it cuts completion to a third,
as five-star prompts did in Netflix's in-product test, a like/dislike
scale is better. Choosing the format by this measure comes
within about 2 percent of the best on average in our experiments.

\item \textbf{Information should be measured at the cutoff.}
The information in a rating depends on the item's value, so we must
choose where to measure it. Measuring it at the cutoff reproduces
the ordinal decision loss within about 7 percent for 1 to 100
responses per item. Averaging it over the values the item might take
tracks the estimation error better but can misstate the decision loss by
about half. Which summary is right thus depends on what the
ratings are used for. With an independent prior and the same large
number of responses about every item, the ratio of the proxy's loss to the ordinal loss tends to one if and only if the proxy gives every item a harmonic average of the information at the cutoff, one number per scale computable from the prior.
\end{enumerate}

\subsection{Related Literature}
\label{subsec:lit}
\paragraph{Selecting top alternatives under a measurement budget.}
Pure exploration in multi-armed bandits seeks the best
arm or the top $K$ arms
\citep{KalyanakrishnanEtAl,GabillonEtAl,KaufmannEtAl}. In ranking and
selection, optimal computing budget allocation determines how many
simulation replications each alternative receives, and a variant
selects the top $K$ alternatives \citep{ChenHeFuLee2008}. One-stage and
two-stage procedures plan sampling in advance but select a single best
alternative \citep{GuptaMiescke1996,ChickInoue2001}. Uncertainty sampling in active learning asks about the
cases the model is most uncertain about \citep{Settles2009},
regardless of whether the answer could change the selection
(Section~\ref{subsec:marginal-value}). Value-of-information and
knowledge-gradient methods choose measurements by expected
improvement in the decision \citep{FrazierPowellDayanik}, and
evaluate several jointly when the value of information is not concave \citep{FrazierPowell2010}. Our envelope rule adapts this block valuation to a top-$K$ plan made before any response arrives, and to each wave of a questionnaire sent in waves (Section~\ref{subsec:num-waves}).

The studies above take the precision of each measurement as given. As with binary feedback
\citep{WangWangPowell2016}, the information in an ordinal rating
depends on the item's unknown value. We show which single summary of it
makes the Gaussian proxy exact for the top-$K$ decision with many
responses (Proposition~\ref{prop:kappa-star}). Because this summary is
the same for every item, formats can be compared by one number and
chosen separately from the allocation
(Proposition~\ref{prop:eis-ranking}).

\paragraph{Measuring information for decisions.}
Optimal design and the design of choice experiments choose questions to
estimate the model parameters precisely
\citep{ChalonerVerdinelli1995,SandorWedel2001}. Gaussian approximations
of ordinal models are built to match the posterior
\citep{ChuGhahramani2005}. Decision-focused learning instead trains
prediction models to minimize the resulting decision cost
\citep{ElmachtoubGrigas2022}. Computerized classification testing needs
a pass-or-fail decision, not an ability estimate, so it asks the
questions most informative at the pass mark
\citep{SprayReckase1994,Eggen1999,Thompson2009}. We apply this idea to
our Gaussian proxy and measure the information in a rating at the
selection cutoff (Section~\ref{subsec:where-info}).

\paragraph{Ordinal elicitation, survey burden, and feedback formats.}
Ordinal regression models such as the ordered probit link ratings to
unobserved values \citep{McCullagh}. In marketing, adaptive conjoint designs
choose each question based on the answers so far
\citep{ToubiaHauserSimester2004,SaureVielma2019}. Longer or more demanding
questionnaires reduce participation and can lower response quality
\citep{Krosnick1991,Galesic2009,Callegaro2015}. Studies of rating
scales examine how the number and labels of categories affect
reliability and validity \citep{PrestonColman2000,KrosnickPresser2010}.
We combine questionnaire length, completion rate, and information per
rating at the cutoff into the expected information per customer asked
(Section~\ref{sec:format}).

\paragraph{Subscription and curated retail.}
Customers who subscribe to curated services delegate the choice to the
platform and bear the risk of receiving products they dislike
\citep{BischofBoettgerRudolph2020}. Operations research on subscription
and curated retail studies which products to put in a box
\citep{BernsteinGuo2023} and which to offer while learning demand
\citep{CaroGallien2007,SaureZeevi2013,BernsteinModaresiSaure2019,KallusUdell2020}.
In these models the platform knows what customers like or learns it
from sales. We study how the
platform should gather this knowledge from customer feedback.

\section{Model and Optimization Problem}
\label{sec:model}

We model one round of selection, in which the platform sends a single
questionnaire and uses the ratings that come back to fill the box with
candidate items.

\subsection{Item values}
\label{subsec:items}\label{subsec:unknown}
The platform fills a box with \(K\) of \(L\) candidate items
for a target customer population, where \(1\le K<L\). The candidate items
\(\mathcal I:=\{1,\ldots,L\}\) are those that could go in the box, such as
items in stock that meet dietary restrictions. The box is a
subset \(B\subseteq\mathcal I\) with \(|B|=K\).

A customer's utility from an item measures how much the customer likes
it. For a customer drawn from the population, we write the utility from
item \(i\) as
\begin{equation}\label{eq:utility}
u_i=w_i+\varepsilon_i,
\qquad
\varepsilon_i\sim\mathcal N(0,1).
\end{equation}
Here \(w_i\) is the average utility of item \(i\) across the population,
which we call the value of item \(i\), and \(\varepsilon_i\) is the
customer's deviation from that average. Deviations are independent
across items and across customers. We set their variance to one because
ratings do not identify the scale of utility. Customers differ only in
these deviations, so every rating is informative about the same
vector \(\w=(w_1,\ldots,w_L)\).

We assume that a customer's utility from a box \(B\) is the sum of
its item utilities, \(\sum_{i\in B}u_i\). The expected utility of
\(B\) for a customer drawn from the population is then
\(\sum_{i\in B}w_i\), which the platform seeks to maximize. A platform
that knew \(\w\) would choose the \(K\) items with the largest values.

The platform does not know \(\w\). It holds a Gaussian prior
\begin{equation}\label{eq:prior}
\w\sim\mathcal N(\bmu,\bSigma),
\end{equation}
with mean \(\bmu\in\R^L\) and positive-definite covariance
\(\bSigma\in\R^{L\times L}\). The prior mean reflects information
available before feedback, for example product attributes or
experience with similar items. The off-diagonal entries of \(\bSigma\) let a
rating of one item update beliefs about related items, such as
recipes from the same cuisine or garments of the same style.

\subsection{Ordinal customer feedback}
\label{subsec:feedback}\label{subsec:estimator}
Customers do not report utilities. When asked about item \(i\), a
customer gives a rating on an ordered scale with \(S\) categories,
\(R_i\in\{1,\ldots,S\}\). For example, \(S=2\) is a binary
like/dislike scale and \(S=5\) a five-point scale. Under an
ordered-probit model, the customer reports category \(s\) when the
utility falls between two thresholds,
\[
R_i=s
\quad\Longleftrightarrow\quad
\theta_{s-1}<u_i\le\theta_s,
\qquad
-\infty=\theta_0<\theta_1<\cdots<\theta_S=+\infty .
\]
Because \(\varepsilon_i\) is standard normal, the rating probabilities
are
\begin{equation}\label{eq:ordered-probit}
\Pr(R_i=s\mid w_i)
=\Phi(\theta_s-w_i)-\Phi(\theta_{s-1}-w_i),
\qquad s=1,\ldots,S,
\end{equation}
where \(\Phi\) is the standard normal distribution function. We refer
to each completed rating as a response. Responses are independent
given \(\w\). We treat the thresholds as known; they can be estimated
from past ratings on the same scale.

A single rating shows only which interval contains the utility. We
measure how precisely one rating pins down an item's value \(w\) by its
Fisher information. With \(\phi\) the standard normal density and
\(\phi(\pm\infty)=0\), a rating on scale \(S\) has information
\begin{equation}\label{eq:fisher-info}
I_S(w)
=\sum_{s=1}^{S}
\frac{\big[\phi(\theta_{s-1}-w)-\phi(\theta_s-w)\big]^2}
{\Phi(\theta_s-w)-\Phi(\theta_{s-1}-w)} .
\end{equation}
Observing the utility itself would give information equal to one, so
\(I_S(w)<1\). The information a rating provides depends on where \(w\) lies
relative to the thresholds. It is large when \(w\) is near a threshold, where customers
split between adjacent categories and the rating probabilities change
quickly with \(w\). For \(w\) far above or below all thresholds, nearly every
customer gives the same rating and \(I_S(w)\) is small. Splitting a
category in two makes a rating more informative, and the positions of
the thresholds also matter (Section~\ref{subsec:eis-calibration}).
Since \(I_S(w)\) depends on the unknown \(w\), Section~\ref{sec:proxy}
summarizes it by one number per item, \(\kappa_i(S)\).

\subsection{Feedback collection and allocation}
\label{subsec:collection}\label{subsec:design-problem}
The platform collects feedback through a questionnaire. A feedback
format consists of the number \(Q\) of items each customer is asked
to rate and the response scale \(S\) of Section~\ref{subsec:feedback}. The expected fraction of
customers shown the questionnaire who complete it is the completion
rate \(\alpha(Q,S)\in[0,1]\), which is lower for more demanding formats. The size of the
target population sets the number \(T\) of customers shown the
questionnaire. The expected number of completed
responses is then \(N_{\mathrm{tot}}(Q,S)=T\,Q\,\alpha(Q,S)\). The
questionnaire is sent once, for example by email after a delivery, so
the platform fixes the whole design before any answer arrives.

The platform also chooses which items each customer rates, and
thus how many responses each item receives. Let \(N_i\) denote the planned
number of completed responses about item \(i\). The entries of the
allocation \(\N=(N_1,\ldots,N_L)\) sum to at most
\(N_{\mathrm{tot}}(Q,S)\). With many customers,
realized counts stay close to their planned values, so we use the
continuous relaxation that treats each \(N_i\) as a nonnegative real
number. We also drop the limit of one response per item from each customer who
completes the questionnaire. Imposing it matters little for the chosen
format (Section~\ref{subsec:num-format}).

\subsection{The top-\texorpdfstring{$K$}{K} decision and decision loss}
\label{subsec:decision-loss}\label{subsec:retention}
After collecting the ratings \(\cD(\N;S)\) under scale \(S\) and
allocation \(\N\), the platform estimates the values by the
posterior mean \(\hat{\w}(\N;S):=\E\big[\w\mid\cD(\N;S)\big]\) and
selects the \(K\) items with the highest posterior means,
\[
\hat B(\N;S)\in\arg\max_{B\subseteq\mathcal I,\ |B|=K}\
\sum_{i\in B}\hat w_i(\N;S).
\]
Given the data, any set \(B\) has expected value
\(\sum_{i\in B}\hat w_i(\N;S)\), so plugging in the posterior means for
\(\w\) is optimal. The benchmark
is an oracle that knows \(\w\) and selects
\[
B^*(\w)\in\arg\max_{B\subseteq\mathcal I,\ |B|=K}\ \sum_{i\in B}w_i .
\]

The sum of the \(K\) largest entries of a vector \(\mathbf x\in\R^L\) is
\(V_K(\mathbf x):=\sum_{r=1}^{K}x_{(r)}\), where
\(x_{(1)}\ge\cdots\ge x_{(L)}\) are the entries of \(\mathbf x\) in
decreasing order. Then \(V_K(\w)\) is the total value of the oracle
selection. We measure a feedback design by its decision loss, the
expected amount by which the total value of the plug-in selection falls
short of the oracle's,
\begin{equation}\label{eq:def-delta}
\Delta_{\mathrm{ord}}(\N;S)
:=\E\Big[\sum_{i\in B^*(\w)}w_i\Big]
-\E\Big[\sum_{i\in\hat B(\N;S)}w_i\Big]
=\E\big[V_K(\w)\big]-\E\big[V_K(\hat{\w}(\N;S))\big].
\end{equation}
The subscript refers to the ordinal ratings of Section~\ref{subsec:feedback}.
The second equality holds because, given the data, the plug-in
selection has expected value \(V_K(\hat{\w}(\N;S))\).

\subsection{The joint feedback-design problem}
\label{subsec:joint-problem}
For a given number \(T\) of customers shown the questionnaire, the
platform chooses the format and the allocation to minimize decision
loss,
\begin{equation}\label{eq:joint-problem}
\begin{aligned}
\min_{Q,\,S,\,\N}\quad & \Delta_{\mathrm{ord}}(\N;S)\\
\text{s.t.}\quad & \sum_{i=1}^{L}N_i\le T\,Q\,\alpha(Q,S),
\qquad N_i\ge0\quad(i=1,\ldots,L).
\end{aligned}
\end{equation}
The format sets both the response budget
\(T\,Q\,\alpha(Q,S)\) and, through the scale \(S\), how informative each
response is. Section~\ref{subsec:num-robust} also considers
sending the questionnaire in waves and replanning between
them.

The decision loss has no closed form under ordinal ratings, so we
replace the ratings by a Gaussian proxy (Section~\ref{sec:proxy}).
Because the best allocation depends on the format, we plan each
format's allocation under the proxy (Section~\ref{sec:allocation}) and
then choose among formats (Section~\ref{sec:format}).

\section{An Information-Equivalent Gaussian Proxy}
\label{sec:proxy}\label{subsec:proxy}

Under ordinal ratings, the posterior is not normal, so the decision loss
of a design must be simulated. This is slow and gives no formula for how
the loss depends on the allocation or the format. The proxy is a Gaussian model
whose signals carry the same information as the ratings. Its decision
loss depends on the design only through a covariance matrix. After defining
the proxy (Section~\ref{subsec:proxy-def}), we choose where to measure a
rating's information (\ref{subsec:where-info}) and check the proxy against
the ordinal model (\ref{subsec:proxy-validation}).

\subsection{The proxy}
\label{subsec:proxy-def}
The proxy replaces each ordinal response about item \(i\) by a
normal signal centered at \(w_i\). We summarize the information in one response under scale \(S\) by a
number \(\kappa_i(S)\), chosen in Section~\ref{subsec:where-info}. With variance
\(1/\kappa_i(S)\), the signal is as informative as the response, and in
this sense information-equivalent.

After \(N_i\) signals about each item \(i\), standard Gaussian updating
gives a normal posterior. Its covariance measures the
uncertainty about \(\w\) that remains, and we call it
the residual covariance,
\begin{equation}\label{eq:residual-cov}
A(\N;S):=\Big(\bSigma^{-1}+\diag\big(N_1\kappa_1(S),\ldots,
N_L\kappa_L(S)\big)\Big)^{-1}.
\end{equation}
The residual covariance depends on the allocation and on how
informative each response is, but not on the responses received. Formula~\eqref{eq:residual-cov} applies to any nonnegative real
allocation, as the continuous relaxation of Section~\ref{subsec:collection}
requires.

The allocation is planned before any response arrives, when the
posterior mean is not yet known. Under the proxy, this posterior mean is
normally distributed,
\begin{equation}\label{eq:proxy-postmean}
\tilde{\w}(\N;S)\sim\mathcal N\big(\bmu,\ \bSigma-A(\N;S)\big).
\end{equation}
Its covariance is the part of the prior uncertainty that the feedback
removes. Without feedback,
\(\tilde{\w}\) equals \(\bmu\); as all \(N_i\) grow, its distribution
approaches that of \(\w\).

Replacing \(\hat{\w}(\N;S)\) by \(\tilde{\w}(\N;S)\) in the last
expression of \eqref{eq:def-delta} gives the proxy decision loss
\begin{equation}\label{eq:delta-proxy}
\Delta(\N;S):=\E\big[V_K(\w)\big]-\E\big[V_K(\tilde{\w}(\N;S))\big].
\end{equation}
By \eqref{eq:proxy-postmean}, \(\Delta(\N;S)\) depends on the design only
through \(A(\N;S)\) and is therefore much easier to evaluate and
optimize than \(\Delta_{\mathrm{ord}}(\N;S)\). When the scale \(S\) is
fixed, we write \(A(\N)\) and \(\Delta(\N)\).

\subsection{Where to measure the information}
\label{subsec:where-info}
The proxy uses one number for the information in a rating, because a
Gaussian signal is equally informative at every value. By
\eqref{eq:fisher-info}, the information in an ordinal rating instead varies
with the item's value, so the number depends on where it is measured.

We compare two choices for the number and adopt the second.
\begin{enumerate}
\item \emph{Per-item prior average.} Average the information over the
item's prior,
\[
\kappa_i^{\mathrm{avg}}(S):=\E\big[I_S(w_i)\big],\qquad
w_i\sim\mathcal N(\mu_i,\Sigma_{ii}),
\]
as in the Fisher information gain criterion of Bayesian design
\citep{Overstall2022}.
\item \emph{Information at the cutoff.} Give every item the information of a
rating at the cutoff between selected and
unselected items, averaged over where the cutoff is likely to fall,
\begin{equation}\label{eq:kappa}
\kappa(S):=\frac{\E\big[\sum_i\psi_i\big]}
{\E\big[\sum_i\psi_i/I_S(\zeta_i)\big]}.
\end{equation}
\end{enumerate}

Item \(i\) belongs to the best
\(K\) exactly when its value exceeds the \(K\)th largest value among the
other items, the cutoff \(\zeta_i\) that item \(i\) faces.
More generally, for any vector \(\mathbf x\) of values or estimates,
\(\zeta_i(\mathbf x):=\big(\mathbf x_{-i}\big)_{(K)}\) denotes the \(K\)th
largest entry of \(\mathbf x\) without item \(i\). We measure how likely
item \(i\) is to be near its cutoff by the density of its value
there,
\begin{equation}\label{eq:cutoff-faced}
\zeta_i:=\zeta_i(\w),
\qquad
\psi_i:=f_i\big(\zeta_i\mid\w_{-i}\big),
\end{equation}
where \(f_i(\cdot\mid\w_{-i})\) is the prior density of \(w_i\) given the
other values \(\w_{-i}\); under an independent prior it does not
depend on them. Both quantities are random, because the other
values are unknown. Formula~\eqref{eq:kappa} is a harmonic average. It averages
\(1/I_S(\zeta_i)\) with weight \(\psi_i\) and inverts the result, since
the decision loss scales with the residual variance, one over the information. Like the prior average, \(\kappa(S)\) uses
only the prior and can be computed before any response arrives.

We measure the information at the cutoff because the platform estimates
values only to select items. An estimation error is costly only when it
moves an item across its cutoff, so the decision loss depends on the
information at values near the cutoff. Because the prior average also counts
values far from the cutoff, where a rating's information can be very
different, it can be too high or too low for items near the
cutoff. In classification
testing, questions are likewise chosen to be most informative at the
pass mark
\citep{SprayReckase1994,Eggen1999,Thompson2009}.
Proposition~\ref{prop:kappa-star} justifies \(\kappa(S)\) for large
budgets.

\begin{proposition}[Where the information should be measured]
\label{prop:kappa-star}
Suppose the prior is independent across items, and that the inverse
prior variances of the \(\min(K,L-K)+1\) items with the largest prior
variances sum to more than one. Let the proxy give item \(i\) the
information \(\kappa_i\) per response, and suppose every item receives
the same number of responses \(n\). As \(n\to\infty\), the following
hold.
\begin{enumerate}
\item[(i)] \(n\,\Delta_{\mathrm{ord}}(\N;S)\to\frac12\,
\E\big[\sum_i\psi_i/I_S(\zeta_i)\big]\) and
\(n\,\Delta(\N;S)\to\frac12\,\E\big[\sum_i\psi_i/\kappa_i\big]\).
\item[(ii)] A proxy that gives every item the same information is exact
in the limit, that is, its decision loss divided by the ordinal loss
tends to one, if and only if that information equals \(\kappa(S)\).
\end{enumerate}
\end{proposition}

The two limits in part (i) differ only in the information per
response. In both, each item enters through the density of its value at its
cutoff, where a response has information \(I_S(\zeta_i)\) in the ordinal
model and \(\kappa_i\) in the proxy. Part (ii)
follows because \(\kappa(S)\) in \eqref{eq:kappa} averages
\(I_S(\zeta_i)\) with the weights \(\psi_i\) common to both
limits. Dividing the two limits for any choice of
\(\kappa_i\) gives
\begin{equation}\label{eq:kappa-ratio}
\frac{\Delta(\N;S)}{\Delta_{\mathrm{ord}}(\N;S)}\;\to\;
\frac{\E\big[\sum_i\psi_i/\kappa_i\big]}{\E\big[\sum_i\psi_i/I_S(\zeta_i)\big]},
\end{equation}
which Section~\ref{subsec:proxy-validation} uses to predict the error of
the prior average.

By part (ii), we set \(\kappa_i(S)=\kappa(S)\) for every item and write
\(\kappa(S)\) from now on, or \(\kappa\) when the scale is fixed.
Corollary~\ref{cor:p1-shares} in Appendix~\ref{app:proofs-proxy} extends
part (i) to any allocation that gives each item a fixed positive share
of the responses. The appendix also shows that an item-level version of
\(\kappa(S)\) would differ from it by a few percent, since all items
face about the same cutoff. The common value lets
Section~\ref{sec:format} compare formats by one number.

\subsection{How well the proxy reproduces the ordinal model}
\label{subsec:proxy-validation}
The proxy can err for two reasons. It gives every rating of an item the
same information, and it treats the platform's posterior as normal,
which it is not when ratings are few. We compare the proxy with the exact ordinal model under each of the two
choices of Section~\ref{subsec:where-info}, down to one response per
item. With the information at the cutoff, the proxy reproduces the decision
loss at every budget. The per-item prior average predicts the estimation
error better but can misstate the decision loss by almost 60 percent.

The platform selects \(K=6\) of \(L=60\) candidate items, and the prior
is independent with unit variances. The binary
scale has one threshold at zero, and the five-point scale has thresholds
at \(-1\), \(-0.2\), \(0.2\), and \(1\). We consider three configurations of the prior means. In the random configuration they are
drawn from a standard normal distribution. In the separated and crowded configurations, neighboring prior means
around the \(K\)th place are 0.40 and 0.055 apart. The crowded
configuration thus has many items near the cutoff. All items receive the same number of responses, from 1 to 3,000, and the
tables report 10, 100, and 1,000.
We compute the ordinal posterior exactly by numerical integration. The
proxy's normal signals reuse the noise drawn for the ratings, which
reduces the simulation noise in the difference between the two models. Each result averages 2,000 replications.

\begin{table}[t]
\centering
\caption{Ratio of the estimation error that the proxy predicts to the
mean squared error of the ordinal posterior means, under a uniform
allocation. Bold marks the ratio closer to one in each pair.}
\label{tab:proxy-posterior}
\small
\begin{tabular}{@{}ll*{3}{w{c}{2.6em}}@{\hspace{14pt}}*{3}{w{c}{2.9em}}@{}}
\toprule
 & & \multicolumn{3}{c}{Per-item prior average} & \multicolumn{3}{c}{Information at the cutoff} \\
\cmidrule(r{8pt}){3-5}\cmidrule(l){6-8}
\multicolumn{2}{@{}l}{Responses per item} & 10 & 100 & 1,000 & 10 & 100 & 1,000 \\
\midrule
Random & Binary & \textbf{0.81} & 0.48 & 0.24 & 1.46 & \textbf{1.06} & \textbf{0.55} \\
 & Five-point & 0.87 & 0.69 & 0.52 & \textbf{1.11} & \textbf{0.90} & \textbf{0.68} \\
\addlinespace
Separated & Binary & \textbf{0.77} & \textbf{0.40} & \textbf{0.18} & 0.50 & 0.23 & 0.10 \\
 & Five-point & \textbf{0.82} & \textbf{0.59} & \textbf{0.43} & 0.61 & 0.42 & 0.31 \\
\addlinespace
Crowded & Binary & \textbf{0.77} & \textbf{0.43} & \textbf{0.21} & 0.53 & 0.27 & 0.13 \\
 & Five-point & \textbf{0.84} & \textbf{0.64} & \textbf{0.51} & 0.66 & 0.49 & 0.39 \\
\bottomrule
\end{tabular}
\end{table}

\begin{table}[t]
\centering
\caption{Ratio of the proxy decision loss to the ordinal decision loss
under a uniform allocation. Bold marks the ratio closer to one in each
pair, and Limit is the large-budget value from \eqref{eq:kappa-ratio}.}
\label{tab:proxy-loss}
\small
\begin{tabular}{@{}ll*{4}{w{c}{2.6em}}@{\hspace{14pt}}*{3}{w{c}{2.9em}}@{}}
\toprule
 & & \multicolumn{4}{c}{Per-item prior average} & \multicolumn{3}{c}{Information at the cutoff} \\
\cmidrule(r{8pt}){3-6}\cmidrule(l){7-9}
\multicolumn{2}{@{}l}{Responses per item} & 10 & 100 & 1,000 & Limit & 10 & 100 & 1,000 \\
\midrule
Random & Binary & 0.58 & 0.49 & 0.48 & 0.48 & \textbf{0.99} & \textbf{1.01} & \textbf{0.95} \\
 & Five-point & 0.81 & 0.81 & 0.83 & 0.80 & \textbf{0.98} & \textbf{1.02} & \textbf{1.02} \\
\addlinespace
Separated & Binary & 1.26 & 1.37 & 1.38 & 1.41 & \textbf{0.98} & \textbf{1.01} & \textbf{0.97} \\
 & Five-point & 1.21 & 1.27 & 1.33 & 1.23 & \textbf{1.04} & \textbf{1.03} & \textbf{1.13} \\
\addlinespace
Crowded & Binary & 1.18 & 1.37 & 1.32 & 1.31 & \textbf{0.97} & \textbf{1.06} & \textbf{1.03} \\
 & Five-point & 1.16 & 1.11 & 1.15 & 1.18 & \textbf{1.02} & \textbf{0.96} & \textbf{0.97} \\
\midrule
\multicolumn{2}{@{}l}{Largest standard error} & 0.03 & 0.05 & 0.09 &  & 0.02 & 0.04 & 0.08 \\
\bottomrule
\end{tabular}
\end{table}

Table~\ref{tab:proxy-posterior} compares the estimation error, the mean
squared error of the ordinal posterior means, with the proxy's average
residual variance. The
per-item prior average comes closer in most settings, but both choices
understate the error once items have many responses. The error, averaged over all items, is dominated by items far from every
threshold, whose ratings provide almost no information. The prior average partly reflects this, but any fixed information
per rating makes the predicted error fall too fast.

For the decision loss (Table~\ref{tab:proxy-loss}), the ratio is closer
to one under the information at the cutoff in every case. It is within 6 percent of one for 10 to 100 responses per item and
0.92 to 1.13 across six budgets up to 3,000, with no
consistent sign (Appendix~\ref{app:proofs-proxy}). With 10 responses per
item, \(n\Delta_{\mathrm{ord}}\) is only 66 to 88 percent of its limit,
yet the ratio is 0.97 to 1.04.
Treating the posterior as normal costs little even with 1, 2, or 5
responses per item, where the ratio is 0.93 to 1.01.

Under the per-item prior average, the direction of the error depends on
where the cutoff lies relative to the thresholds. In the random configuration, items near the cutoff have values well
above the thresholds, where a rating is less informative than the prior
average implies. The proxy then understates the loss by about 50 percent
under the binary scale and 20 percent under the five-point scale. In the other two, the cutoff lies near the thresholds,
and the proxy overstates the loss by 11 to 57 percent across budgets.
Proposition~\ref{prop:kappa-star} predicts these errors from the prior
alone. At 1,000 responses per item, the observed ratios are within 0.10 of the
Limit column of Table~\ref{tab:proxy-loss}.

With the information at the cutoff, the proxy stays accurate in the other
settings of Sections~\ref{sec:allocation} to~\ref{sec:numerics}. The ratio is 0.95 to 1.06 for the envelope and square-root rules (Section~\ref{subsec:num-allocation}), and 0.89 to 1.12 with prior variances 0.3 and 0.1, or when items with prior means near the \(K\)th largest get more responses.
Appendix~\ref{app:proofs-proxy} repeats the comparison with correlated
items.

The two choices differ most in ranking formats when the cutoff
lies far from the rating thresholds. In the random configuration, when completion drops steeply for more
demanding formats (Section~\ref{subsec:num-format}), the proxy and
ordinal rankings of five formats have rank correlation
0.81 to 0.95 under the information at the cutoff and 0.57 to 0.67 under
the prior average. The proxy is weakest within a single replication. For binary ratings
and 10 responses per item, the two models share 80 percent of selected
items on average, though their expected losses agree. Planning uses
only expected losses and rankings. In
Section~\ref{subsec:proxy-ranking}, proxy and ordinal rankings of
allocations and formats agree about as closely as two independent
ordinal simulations. The ordinal decision loss remains the objective.
Sections~\ref{sec:allocation} and~\ref{sec:format} derive their results
under the proxy, and Section~\ref{sec:numerics} evaluates the resulting
rules under the ordinal model.

\FloatBarrier

\section{Allocating Feedback Across Items}
\label{sec:allocation}\label{sec:alloc_opt}

We take the format \((Q,S)\) as given, which fixes the response budget
\(N_{\mathrm{tot}}=T\,Q\,\alpha(Q,S)\) and the information per response
\(\kappa=\kappa(S)\). The platform then chooses how to distribute
the \(N_{\mathrm{tot}}\) responses across the \(L\) items,
\begin{equation}\label{eq:alloc-problem}
\min_{\N\ge 0}\ \Delta(\N)
\qquad\text{subject to}\qquad
\sum_{i=1}^{L}N_i\le N_{\mathrm{tot}} .
\end{equation}
The questionnaire is sent once, so the allocation is planned from the
prior. The results below also hold when the information per
response differs across items (Appendix~\ref{app:proofs-alloc}).

Responses are worth most for items
whose place in the selection is unclear. For an item far from the cutoff, the value
of responses grows slowly and then faster, so a plan
should consider blocks of responses.

Several steps use known tools, namely the heat-equation identity, the knowledge gradient, budget allocation across activities with S-shaped returns, and stratified sampling. Their use to plan from the prior for a top-\(K\) selection is new. It weights each item by how likely it is to be near the cutoff and gives a square-root law for the gain from planning.

\subsection{Why the top-\texorpdfstring{$K$}{K} boundary matters}
\label{subsec:cutoff-logic}
Section~\ref{subsec:where-info} argued that only errors near the cutoff
matter. Write
\(w_{(1)}\ge\cdots\ge w_{(L)}\) for the true item values in decreasing
order, and let \(\bar w:=\big(w_{(K)}+w_{(K+1)}\big)/2\) be the midpoint
between the last selected item and the first excluded one.

\begin{lemma}[Stability of the selection]
\label{lem:cutoff}\label{lem:wrong-swap-band}
Fix \(\w\) and an estimate \(\hat{\w}\). If each item's error
is smaller than the distance of its value from the midpoint,
\[
\big|\hat w_i-w_i\big|<\big|w_i-\bar w\big|\qquad\text{for every } i,
\]
then the \(K\) items with the highest estimates are the \(K\) items with
the highest values. In particular, this holds when every error is
smaller than half the gap \(w_{(K)}-w_{(K+1)}\).
\end{lemma}

\begin{figure}[tb]
\centering
\includegraphics{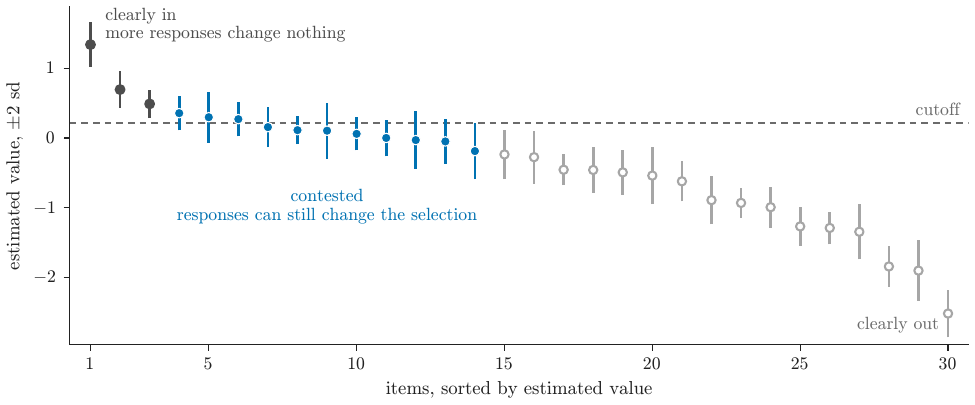}
\caption{The cutoff mechanism. Items sorted by their estimated values,
with \(\pm 2\) standard-deviation intervals.}
\label{fig:cutoff-mechanism}
\end{figure}

An item far from the cutoff can thus be estimated poorly without
changing the selection. Figure~\ref{fig:cutoff-mechanism}
shows the prior means, which are the estimates before any
feedback. We call the \(K\)th largest prior mean the prior cutoff.
Items whose intervals straddle it are contested; responses
about them can change the selection.

\subsection{How the decision loss splits across items}
\label{subsec:loss-representation}\label{subsec:bound-N}
Unless stated otherwise, we assume that the prior is
independent across items. Let \(\tau_i\) be the prior precision of item
\(i\). After \(N_i\) responses, item \(i\) has residual variance
\(a_i(\N):=1/(\tau_i+\kappa N_i)\), the \(i\)th diagonal entry of the
residual covariance \(A(\N)\). Write \(X:=\tilde{\w}(\N)\) for the
estimates that the platform will hold, viewed before responses
arrive, so that \(X\sim\mathcal N\big(\bmu,\ \bSigma-A(\N)\big)\) by
\eqref{eq:proxy-postmean}.

More responses about item \(i\) change the selection only
when its estimate \(X_i\) is close to the cutoff \(\zeta_i(X)\) set by the
other estimates. There, residual uncertainty causes near misses, in which the platform
selects an item whose value is just below the cutoff or leaves out one
just above it. Their expected cost is about half the density of the estimate at the
cutoff, times the residual variance.

\begin{proposition}[Decision loss by item]
\label{prop:loss-representation}\label{prop:gaussian-gap}
The proxy decision loss \(\Delta(\N)\) is nonnegative for every prior
covariance and depends on the allocation only through the residual
covariance \(A(\N)\). If the prior is independent, then
\[
\Delta(\N)=\sum_{i=1}^{L}c_i(\N)\,a_i(\N),
\qquad
c_i(\N):=\frac12\int_0^1\E\big[f_{i,t}\big(\zeta_i(X_t)\big)\big]\,dt\ \ge 0 ,
\]
where \(X_t\sim\mathcal N\big(\bmu,\ \bSigma-(1-t)A(\N)\big)\), whose law
moves from that of the estimates \(X\) at \(t=0\) to that of the true
values at \(t=1\), and \(f_{i,t}\) is the density of its \(i\)th entry.
\end{proposition}

We call \(c_i(\N)\) the \emph{cutoff weight} of item \(i\). It is the
near-miss cost per unit of residual variance and
reflects how likely the item is to be near the cutoff.

\subsection{The value of one more response}
\label{subsec:marginal-value}\label{subsec:alloc-local-value}
The value of one more response about an item is the rate at which the
decision loss falls as the item receives more responses. With the
numbers of responses treated as continuous
(Section~\ref{subsec:collection}), this rate is
\[
\mathrm{MV}_j(\N):=-\frac{\partial\Delta(\N)}{\partial N_j},
\]
which we call the marginal value of responses to item \(j\).

More responses spread the estimates \(X\) further from the prior means.
The wider spread raises the expected value of the selection only
through items whose estimates can cross their cutoffs. The \emph{local cutoff weight}
\(b_j(\N)\) measures how likely the estimate of item \(j\) is to be
near its cutoff under the planned allocation. Formally, it is half
the density of the gap \(X_j-\zeta_j(X)\) at zero. The cutoff weight \(c_j(\N)\) averages this density along the path
from the estimates to the true values, and \(b_j(\N)\) is its value at
the start. With many responses about every item, the two weights
approach the same limit (Section~\ref{subsec:sqrt-law}).

\begin{proposition}[Marginal value of responses]
\label{prop:marginal-value}\label{prop:local-marginal}
Under the Gaussian proxy and an independent prior,
\begin{equation}\label{eq:priority-decomposition}
\mathrm{MV}_j(\N)
=\underbrace{b_j(\N)}_{\text{local cutoff weight}}
\times
\underbrace{\frac{\kappa}{(\tau_j+\kappa N_j)^2}}_{\text{uncertainty one more response removes}} ,
\end{equation}
with the right derivative when \(N_j=0\), provided no other item without
responses has prior mean \(\mu_j\).
\end{proposition}

The marginal value of a response is the item's local cutoff weight
times the residual variance it removes. The second factor falls as the item receives responses, so each
removes less uncertainty than the one before. A rule that asks about the
most uncertain items ranks them by the second factor alone. It ranks items
as marginal value does only when all local cutoff weights
are equal. In the experiments with uneven prior information and with
MovieLens ratings (Sections~\ref{subsec:num-robust}
and~\ref{subsec:num-movielens}), this rule does about as well as
uniform allocation. The first factor also changes with the item's own
responses, and Section~\ref{subsec:exploration} shows that it can make the
marginal value rise.

The rules below ignore correlation between items, at little
cost (Section~\ref{subsec:num-robust}).

\subsection{The value of a block of responses}
\label{subsec:exploration}\label{subsec:block-value}
The marginal value of the next response can understate what several
responses are worth. Fix the plan for the other items and let item \(j\)
receive more responses. The second factor of
\eqref{eq:priority-decomposition} falls steadily, to a quarter each time
the information about the item doubles. The local cutoff weight \(b_j\)
instead depends on the spread of the estimate of item \(j\) around its
prior mean, which widens with the item's responses.
Appendix~\ref{app:proofs-alloc} establishes three properties of \(b_j\).
\begin{enumerate}
\item \emph{It starts near zero for an item far from the cutoff.} With no
responses, the estimate equals the prior mean. If the prior mean lies
far below the cutoff, or far above it, the estimate cannot reach the
cutoff, and \(b_j\) is close to zero.
\item \emph{It then rises quickly.} The chance that the estimate reaches the
cutoff is tiny at first and grows many times over, so \(b_j\) initially
rises much faster than the second factor falls.
\item \emph{It levels off.} The spread of the estimate approaches
the prior uncertainty about the item, so \(b_j\) tends to half the
prior density of the item's value at the cutoff. If the
prior mean lies near the cutoff, \(b_j\) peaks early and then falls
toward this limit.
\end{enumerate}

The marginal value of a distant item therefore rises at first and falls
once \(b_j\) grows more slowly than the second factor falls. For an item
near the cutoff, it falls from the first response.

\begin{figure}[tb]
\centering
\includegraphics{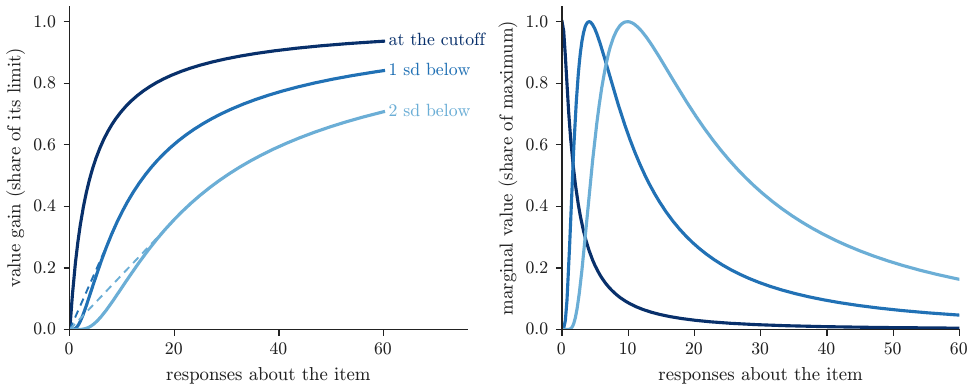}
\caption{Value of responses to items at the prior cutoff and one and two
prior standard deviations below it. Left, the gain over no
responses as a share of its limit, with concave envelopes dashed. Right,
the marginal value as a share of its maximum.}
\label{fig:s-shape}
\end{figure}

The value of a block of responses is the sum of these marginal values.
Item \(j\) is selected exactly when its estimate exceeds
its cutoff \(\zeta_j(X)\). The expected value of the selection
is therefore the expected sum of the \(K\) highest other estimates plus the
expected amount by which the estimate of item \(j\) after \(n\) responses,
\(X_j(n)\), exceeds its cutoff,
\begin{equation}\label{eq:item-value}
\mathcal V_j(n):=\E\big[\big(X_j(n)-\zeta_j(X)\big)^+\big].
\end{equation}
We call \(\mathcal V_j\) the value of item \(j\)'s responses. Its slope
is the marginal value in \eqref{eq:priority-decomposition}, and \(\mathcal V_j\) is
convex where the marginal value rises and concave where it falls. For an item far from the
cutoff, \(\mathcal V_j\) is therefore S-shaped, convex for small numbers
of responses and concave for large ones (Figure~\ref{fig:s-shape}). This
is a form of the nonconcavity of the value of information
\citep{RadnerStiglitz1984,FrazierPowell2010}. Appendix~\ref{app:proofs-alloc}
gives \(\mathcal V_j\) in closed form.

Proposition~\ref{prop:s-shape}, Theorem~1 of \citet{FrazierPowell2010}
in our notation, locates the peak of item \(j\)'s marginal value for a fixed
cutoff \(\zeta\). Write the distance
of the item's prior mean from the cutoff, in prior standard deviations,
as \(z_j:=\sqrt{\tau_j}\,|\mu_j-\zeta|\).

\begin{proposition}[When the value of responses is S-shaped]
\label{prop:s-shape}
Suppose the prior is independent and the cutoff that item \(j\) faces is
fixed at \(\zeta\). The marginal value of responses to item \(j\)
increases in \(N_j\) while \(\kappa N_j<\tau_j\,x(z_j)\) and decreases
afterwards, where
\[
x(z):=\frac{(z^2-1)+\sqrt{(z^2-1)^2+16z^2}}{8}.
\]
Hence \(\mathcal V_j\) is convex below this point and concave above it.
The point grows with the distance from the cutoff, with \(x(0)=0\),
\(x(1)=1/2\), \(x(2)\approx1.44\), and \(x(z)/z^2\to1/4\) as
\(z\to\infty\).
\end{proposition}

The ratio \(\kappa N_j/\tau_j\) is the information in the responses
relative to that in the prior. The marginal value of an item one prior
standard deviation from the cutoff peaks once its responses provide half
as much information as the prior, and for an item two standard
deviations away, about 1.4 times as much. The ratio at the peak grows roughly as
the square of the distance, and a more informative prior delays the peak.
Numerically, the S shape remains when the cutoff varies. Figure~\ref{fig:s-shape} uses
binary ratings, \(L=60\) candidate items with prior means at evenly
spaced normal quantiles and prior variance \(1\), \(K=6\), and ten
responses per other item. The item at the prior cutoff gains
most from its first response. For the items one and two prior standard
deviations below it, the marginal value starts near zero and peaks after
about four and ten responses.

\subsection{Planning the allocation}
\label{subsec:envelope-rule}\label{subsec:guarded}\label{subsec:global-solver}
A natural plan allocates by marginal value. The \emph{marginal-value
rule} starts with no responses and assigns each response to the item for
which one more response lowers the proxy decision loss most, until
the budget is used. It uses this drop because the derivative \(\mathrm{MV}_j\) is zero for
every item at the start. The
knowledge gradient \citep{GuptaMiescke1996,FrazierPowellDayanik2008}
applies the same rule sequentially, as results arrive.

The marginal-value rule is myopic and under-explores items far from the cutoff
(Proposition~\ref{prop:s-shape}). The first response about such an item
is worth little, so the rule asks about it only after every other item's
next response is worth less. Until then, responses go to items near the
cutoff, and the budget may run out first. With prior variance \(0.3\)
and 300 responses per item, the rule leaves most items without a
response and does about 10 percent worse than uniform allocation
(Section~\ref{subsec:num-allocation}).

A minimum share per item covers distant items but spends responses on
items whose place is already clear from the prior
(Section~\ref{subsec:num-allocation}). A better remedy is to value responses in blocks. An item far from the
cutoff is credited with the largest average value per response over a
block of its responses, and receives that block when this average
exceeds the value per response on other items. This replaces
\(\mathcal V_j\) by its \emph{concave envelope}, the smallest concave
function above it (dashed in Figure~\ref{fig:s-shape}). The envelope is
a straight line with this average as slope until it touches
\(\mathcal V_j\), and equals \(\mathcal V_j\) beyond. The
\emph{envelope rule} is the marginal-value rule applied to these
envelopes. The KG(*)
policy of \citet{FrazierPowell2010} uses the same block average in
sequential sampling. Because
the cutoffs depend on the allocation of the other items, the rule
recomputes the curves each time the planned total grows by a fixed
fraction (Algorithm~\ref{alg:envelope}).

\begin{algorithm}[tb]
\caption{Envelope rule}
\label{alg:envelope}
\begin{algorithmic}[1]
\Require prior means and variances, information per response \(\kappa\),
number of items to select \(K\), budget \(N_{\mathrm{tot}}\), refresh fraction \(g\),
minimum information per batch \(\nu\)
\State \(\N\gets\mathbf 0\)
\While{\(\sum_jN_j<N_{\mathrm{tot}}\)}
  \State \(\ell\gets\min\big(N_{\mathrm{tot}}-\sum_jN_j,\ \max(\nu/\kappa,\ g\sum_jN_j)\big)\)
  \State simulate the cutoff \(\zeta_j(X)\) of every item under the plan \(\N\)
  \State for every item \(j\), compute \(\mathcal V_j(N_j+m)\), \(m=0,\ldots,\ell\), and its concave envelope
  \State assign \(\ell\) responses one at a time to the item whose envelope rises most with its next response, breaking ties by a fixed order of the items
\EndWhile
\State \Return \(\N\)
\end{algorithmic}
\end{algorithm}

The envelope rule values each item's responses with the others held
fixed, so we analyze a separable version of the problem. Fix a
reference allocation \(\N^0\) and let the cutoffs keep the distributions
that \(\N^0\) induces. Write \(\mathcal V_j(n;\N^0)\) for the resulting
value of item \(j\)'s responses and
\(\mathcal V(\N;\N^0):=\sum_j\mathcal V_j(N_j;\N^0)\) for their sum. When
only one item's allocation differs from \(\N^0\), this sum and the
expected value of the selection change equally, so they have
the same gradient at \(\N^0\). Proposition~\ref{prop:envelope} adapts a
classical bound \citep{AgraliGeunes2009,UdellBoyd2016} for allocating a
budget across activities with S-shaped returns.

\begin{proposition}[Allocation with S-shaped values]
\label{prop:envelope}
Fix \(\N^0\), and let \(\eta_j\) be the largest gap between
\(\mathcal V_j(\cdot;\N^0)\) and its concave envelope on
\(\{0,\ldots,N_{\mathrm{tot}}\}\). Let \(\hat\N\) assign the
\(N_{\mathrm{tot}}\) responses one at a time to the item whose envelope
rises most with its next response, breaking ties in a fixed order.
Then
\[
\mathcal V(\hat\N;\N^0)\ \ge\ \max_{\N}\mathcal V(\N;\N^0)\;-\;\max_{j}\eta_j ,
\qquad
\eta_j\le\mathcal V_j(N_{\mathrm{tot}};\N^0)-\mathcal V_j(0;\N^0),
\]
where the maximum is over integer allocations with
\(\sum_jN_j=N_{\mathrm{tot}}\).
\end{proposition}

The allocation \(\hat\N\) is the envelope rule's plan for a single batch
of the whole budget, with the cutoffs of \(\N^0\). In the
separable problem, its shortfall is at most one item's gain from the
whole budget, for any number of items. The marginal-value rule has no
such guarantee. In the instances of Section~\ref{sec:numerics}, the
largest gap \(\eta_j\) is about one percent of the decision loss
of uniform allocation with 30 responses per item and under ten percent
with 300. In Algorithm~\ref{alg:envelope}, refreshing the cutoffs between
batches is a heuristic, and its frequency hardly affects the loss
(Section~\ref{subsec:num-allocation}).

The curves \(\mathcal V_j\) and batch sizes depend on responses only
through the information \(\kappa N_j\), so, up to rounding, each item's
planned information does not depend on the scale. Section~\ref{sec:format}
uses this to compare formats.

\subsection{What planning can gain}
\label{subsec:sqrt-law}
With a large budget, the best plan and its gain over uniform allocation
have simple forms. As every item receives more responses, the cutoff
weight of item \(i\) converges to half the average prior density of its value at
the cutoff it faces. We write this limit as \(c_i^\infty:=\tfrac12\E[\psi_i]\),
with \(\psi_i\) as in \eqref{eq:cutoff-faced}.

\begin{proposition}[Square-root allocation]
\label{prop:sqrt-law}
Suppose the prior is independent, let \(\N^\star\) be an optimal
allocation of \(N_{\mathrm{tot}}\) responses, and let \(\N^{\mathrm u}\)
give every item \(N_{\mathrm{tot}}/L\) responses. As
\(N_{\mathrm{tot}}\to\infty\), the share of responses that \(\N^\star\)
gives to item \(i\) converges to
\(\sqrt{c_i^\infty}\big/\sum_{k}\sqrt{c_k^\infty}\), and
\begin{equation}\label{eq:targeting-gain}
\frac{\Delta(\N^\star)}{\Delta(\N^{\mathrm u})}\;\to\;\frac{L_{\mathrm{eff}}}{L},
\qquad
L_{\mathrm{eff}}:=\frac{\big(\sum_i\sqrt{c_i^\infty}\big)^2}{\sum_i c_i^\infty}.
\end{equation}
\end{proposition}

The optimal shares are proportional to the square roots of the cutoff
weights because an item's marginal value falls like \(1/N_i^2\). The same square-root form
gives the optimal allocation of a sample across strata
\citep{Neyman1934,Cochran1977} and appears in the linear-loss procedure of
\citet{ChickInoue2001}. We call \(L_{\mathrm{eff}}\) the effective number of contested items. It
lies between one and \(L\). It equals \(L\) when every item has the same
cutoff weight and \(m\) when \(m\) items share the weight equally and the
others have none. The gain from planning is therefore large when few
items are contested, as under an informative prior. Computed from the
prior before the questionnaire is sent, \(L_{\mathrm{eff}}/L\) is within
0.03 of the envelope rule's loss ratio with 300 responses per item
(Section~\ref{subsec:num-allocation}).

\section{Choosing the Feedback Format}
\label{sec:format}

The questionnaire format determines how many responses the platform
collects and how much information each provides. A longer questionnaire
or a finer scale gives more information per completed questionnaire, but
fewer customers complete it. A format's value also depends on how its
responses are spread across items, so in principle each candidate format
must be evaluated with its own allocation.

Section~\ref{sec:allocation} took the format \((Q,S)\) as given
and planned the allocation. Let \(\pi\) be an
allocation rule, such as the envelope rule, and let \(\N^\pi(Q,S)\) be its
allocation given the budget \(N_{\mathrm{tot}}(Q,S)=T\,Q\,\alpha(Q,S)\)
and the information per response \(\kappa(S)\). The platform then solves
\begin{equation}\label{eq:outer-problem}
\min_{Q,\,S}\ \Delta\big(\N^\pi(Q,S);S\big),
\end{equation}
the proxy form of problem \eqref{eq:joint-problem} with the allocation
set by the rule.

Under the proxy, formats can be compared by one number, the expected
information per customer asked, without planning an allocation
for each. Sections~\ref{subsec:factorized} and~\ref{subsec:eis} define
this number, and Section~\ref{subsec:eis-ranking} shows when it ranks
formats. Section~\ref{subsec:eis-calibration} explains how to estimate it
and choose a format.

\subsection{Factorized information across response scales}
\label{subsec:factorized}
Section~\ref{subsec:where-info} measures the information of a rating at
the cutoff and gives every item the same amount \(\kappa(S)\) per
response under scale \(S\). Fix a baseline scale \(S_0\), such as a
like/dislike scale. A change of scale from \(S_0\) to \(S\) then multiplies the
information about every item by the same factor, the relative information \(\beta(S)\),
\begin{equation}\label{eq:factorized}
\kappa(S)=\beta(S)\,\kappa(S_0),\qquad \beta(S_0)=1 .
\end{equation}
If the factor
differed across items, a format could not be judged apart from its
allocation.

One response under scale \(S\) is worth \(\beta(S)\) responses under the
baseline scale. If item \(i\) receives \(N_i\) responses under scale
\(S\), we call \(N_i':=\beta(S)\,N_i\) its baseline-equivalent responses.
Because \(N_i\kappa(S)=N_i'\kappa(S_0)\), the allocation \(\N\) under scale \(S\) and \(\N'\) under the baseline
scale give the same residual covariance \eqref{eq:residual-cov} and
hence the same proxy decision loss. The budget constraint becomes
\begin{equation}\label{eq:baseline-budget}
\sum_{i=1}^{L}N_i'\ \le\ T\,Q\,\alpha(Q,S)\,\beta(S).
\end{equation}
In baseline-equivalent units, formats differ only in the
budget on the right-hand side.

\subsection{Effective information supply}
\label{subsec:eis}
Dividing the budget in \eqref{eq:baseline-budget} by \(T\) gives the
expected number of baseline-equivalent responses per customer shown the
questionnaire.

\begin{definition}[Effective information supply]
\label{def:eis}
The effective information supply of a feedback format \((Q,S)\) is
\[
\mathrm{EIS}(Q,S):=Q\,\alpha(Q,S)\,\beta(S).
\]
\end{definition}

The term \(Q\,\alpha(Q,S)\) is the expected number of completed
responses per customer asked, and \(\beta(S)\) is the
information in one response relative to the baseline scale. A finer scale raises \(\beta(S)\) and a longer questionnaire
raises \(Q\), but both can lower \(\alpha(Q,S)\). Earlier work compared rating scales by the
preference information a rating provides per second \citep{KluverEtAl2012} and by
the time and effort that each scale demands \citep{SparlingSen2011}. EIS
instead counts information at the cutoff between selected and
unselected items.

\subsection{When EIS ranks formats}
\label{subsec:eis-ranking}
Let \(\Delta^\star(Q,S)\) be the smallest proxy decision loss that the
platform can reach with format \((Q,S)\) and an allocation within the
budget, and let \(\Delta^\pi(Q,S)\) be the loss under the
envelope rule's allocation \(\N^\pi(Q,S)\).

\begin{proposition}[EIS ranks formats]
\label{prop:eis-ranking}\label{prop:eis-equivalence}
Under the Gaussian proxy and the continuous relaxation of
Section~\ref{subsec:collection},
\begin{enumerate}
\item[(i)] There is a nonincreasing function \(\Psi\), the same for all
formats, such that
\[
\Delta^\star(Q,S)=\Psi\big(T\,\mathrm{EIS}(Q,S)\big).
\]
A format with the largest EIS is optimal for the proxy format choice
when the allocation is optimized.
\item[(ii)] Two formats with the same \(T\,\mathrm{EIS}(Q,S)\) receive the
same baseline-equivalent allocation under the envelope rule, so there is
a function \(\Psi_\pi\), the same for all formats, such that
\(\Delta^\pi(Q,S)=\Psi_\pi\big(T\,\mathrm{EIS}(Q,S)\big)\).
\end{enumerate}
\end{proposition}

Part (i), proved in Appendix~\ref{app:proofs-format}, says that under the
best allocation the loss depends on the format only through its EIS, so
ranking formats by EIS is exact. Part (ii) extends this dependence to the
recommended envelope rule of Section~\ref{sec:allocation}, because the
rule counts responses in baseline-equivalent units
(Section~\ref{subsec:envelope-rule}). Under the rule, a
format with larger EIS is better if the loss never rises with
the budget, that is, if \(\Psi_\pi\) is nonincreasing.
In the experiments of
Section~\ref{sec:numerics}, raising the budget lowers the rule's
proxy loss in more than 99 percent of cases.

Under the ordinal model, with whole responses and the limit on
responses per item, choosing the format by EIS comes within 3 percent
of the best format (Section~\ref{subsec:num-format}).

\subsection{Calibrating candidate formats}
\label{subsec:eis-calibration}\label{subsec:eis-practical}
EIS requires two inputs for each candidate format, the completion rate
\(\alpha(Q,S)\) and the relative information \(\beta(S)\). The completion
rate is best estimated by assigning customers to candidate formats at
random and reading completion from exposure and submission logs. The
platform estimates \(\beta(S)\) by fitting the
ordinal model of Section~\ref{subsec:feedback} to ratings collected
under each scale and computing \(\kappa(S)\) from \eqref{eq:kappa},
which averages over where the cutoff may fall. It then ranks
formats by EIS and simulates the ordinal model to check the decision
loss of the leading formats under the envelope rule.

\paragraph{An illustration with public inputs.}
Table~\ref{tab:eis-calibrated} reports EIS for five candidate formats,
with the relative information from MovieLens ratings
\citep{HarperKonstan2016}. We set the four thresholds of the five-point
scale to match the overall shares of one to five stars, and the value of
each of the 200 movies of Section~\ref{subsec:num-movielens} to match
its mean rating. A rating of four or five stars counts as a like,
assuming a like/dislike answer has the threshold between three and four
stars and the same noise as a rating. We treat the values as known,
since each is fit to many ratings. With 10 movies selected, the cutoff
is the tenth-highest value.

At this cutoff, a five-point rating carries information 0.78 and a
like/dislike answer 0.45, so \(\hat\beta(5)=1.73\). The like/dislike answer is less informative because its threshold is
well below the values of the movies near the cutoff, and most customers
like them. The estimate falls from 1.83 for 5 movies to 1.63 for 20 as
the cutoff approaches that threshold. A scale's information thus depends
on where its thresholds lie relative to the cutoff as well as on how
many categories it has.

We know of no public randomized experiment that varies both the length
and the scale of a feedback questionnaire, so the completion rates in
the table are assumptions chosen to span a wide range. We measure the burden of a questionnaire in binary (like/dislike)
questions and let completion fall by the same proportion, about 13
percent, with each unit of burden. The two scenarios differ in what a
finer scale costs. In the survey scenario, a five-point question counts
as 1.25 binary questions. Completion ranges from 0.29 for the
lightest format (three binary questions) to 0.13 for the heaviest (seven
five-point questions). These are the average response rates of online and
in-app surveys in a practitioner report \citep{Malnik2025}. They
describe channels, not formats. In
the in-product scenario, following Netflix's in-product test
\citep{Netflix2017}, a five-point format has one third the completion
rate of a like/dislike format of the same length.

\begin{table}[t]
\centering
\caption{Effective information supply of five candidate formats in two
completion scenarios. Bold marks the largest EIS in each scenario.}
\label{tab:eis-calibrated}
\small
\begin{tabular}{@{}lc*{4}{w{c}{3.1em}}@{}}
\toprule
 & & \multicolumn{2}{c}{Survey scenario} & \multicolumn{2}{c}{In-product scenario} \\
\cmidrule(lr){3-4}\cmidrule(lr){5-6}
Format & \(\hat\beta(S)\) & \(\alpha\) & EIS & \(\alpha\) & EIS \\
\midrule
Binary, \(Q=3\)  & 1.00 & 0.290 & 0.87 & 0.290 & 0.87 \\
Binary, \(Q=5\)  & 1.00 & 0.219 & 1.10 & 0.219 & \textbf{1.10} \\
Five-point, \(Q=3\) & 1.73 & 0.261 & 1.35 & 0.097 & 0.50 \\
Five-point, \(Q=5\) & 1.73 & 0.184 & \textbf{1.59} & 0.073 & 0.63 \\
Five-point, \(Q=7\) & 1.73 & 0.130 & 1.57 & 0.055 & 0.67 \\
\bottomrule
\end{tabular}
\end{table}

The five-point formats have higher EIS in the survey scenario, where
five and seven questions give almost the same value, and lower EIS in
the in-product scenario. To see why, let \(e\) be the burden of one
five-point question in binary questions, and \(r\) the ratio of five-point to
like/dislike completion at the same burden. The survey
scenario has \(e=1.25\) and \(r=1\), and the in-product scenario has
\(e=1\) and \(r=1/3\). Since each unit of burden reduces completion by the same proportion,
the best questionnaire has the same burden for either scale. At that
burden, the five-point scale has \(\beta(5)\,r/e\) times the EIS of the
like/dislike scale, so it is better when \(\beta(5)\) exceeds \(e/r\), the
factor by which it cuts completed responses at equal burden. With
\(\hat\beta(5)=1.73\), the EIS ratio is 1.38 in the survey scenario and 0.58
in the in-product scenario. The assumed rates 0.29 and 0.13 set the best
length but not the choice of scale, because this ratio does not depend
on how fast completion falls with length.

\section{Numerical and Empirical Evidence}
\label{sec:numerics}

We test the results of Sections~\ref{sec:allocation}
and~\ref{sec:format} under the ordinal model and on real ratings.
Section~\ref{subsec:proxy-ranking} checks how well the proxy ranks
designs. We then compare allocation rules (\ref{subsec:num-allocation}),
the items each rule asks about (\ref{subsec:num-exploration}), and
formats (\ref{subsec:num-format}). We relax assumptions such as
a correct prior and a single questionnaire (\ref{subsec:num-robust})
and apply the rules to MovieLens ratings (\ref{subsec:num-movielens}).

\paragraph{Common setup.}
Unless stated otherwise, the synthetic experiments use \(L=60\)
candidate items and select \(K=6\). The prior means follow the random, separated, and
crowded configurations of Section~\ref{subsec:proxy-validation} and have a
standard deviation of 0.7 to 1 across items. Each item value has prior
variance \(0.1\), \(0.3\), or \(1\), so a variance of \(0.1\) describes a
well-informed platform and \(1\) a poorly informed one. Ratings are
binary or five-point, with the thresholds of
Section~\ref{subsec:proxy-validation}. The information per response
\(\kappa(S)\) is the information at the cutoff. Budgets are 10, 30, 100, and 300
responses per item on average. Every allocation is evaluated by its
ordinal decision loss. We draw true values from the prior and simulate
ratings, and the platform selects the items with the highest exact
posterior means. All rules use the
same simulated ratings. The envelope rule re-estimates the distribution of the
cutoffs from 2{,}000 draws each time the planned number of responses
grows by 10 percent.

\subsection{How well the proxy ranks designs}
\label{subsec:proxy-ranking}
Sections~\ref{sec:allocation} and~\ref{sec:format} choose allocations and
formats by comparing designs under the proxy.
Section~\ref{subsec:proxy-validation} showed that the proxy predicts the
loss of a given allocation. Here we check whether it also
orders designs as the ordinal model does. The two orderings agree about
as closely as two independent ordinal simulations do.

Panel~A of Table~\ref{tab:proxy-ranking} compares ten fixed
allocations for each scale and budget. Four favor items near the prior
cutoff to different degrees, one does the opposite, four are random,
and one is uniform. Panel~B compares the five formats of
Section~\ref{subsec:eis-calibration} under mild, moderate, and steep
completion loss (Section~\ref{subsec:num-format}),
with 10 to 300 customers per item, more than the 3 to 30 used there. For 20 sets of prior means per configuration
(prior variance \(1\)), we estimate the ordinal and proxy decision
losses of all designs from the same 400 replications. Each row averages the three
configurations and, for formats, the numbers of customers.

Our first measure is how similarly the two models
order the designs. It is the share of design
pairs that the two models order the same way minus the share they order
the opposite way. This rank correlation, known as Kendall's tau, equals
one when the two orderings coincide. The second is how often the two
models pick the same best design. The third is the regret, defined as
the ordinal loss of the proxy's top design minus that of the best
design, divided by the gap between the best and the worst design. To
gauge simulation noise, the Benchmark columns report the same measures
for two independent 400-replication runs of the ordinal simulation on
the same true values.

For allocations, the rank correlation is 0.89 to 0.93 and the regret
is below 2 percent. For formats, the rank correlation falls from 0.92 under mild to 0.74
under steep completion loss, where several formats perform almost
equally well. In every row, the proxy is within 0.02 of the benchmark in rank
correlation and within 0.7 percentage points in regret. At large budgets, designs favoring items near the cutoff nearly
tie, so the two models, like two ordinal runs, often pick different
best allocations. The results
are similar with 1 to 5 responses per item
(Appendix~\ref{app:numerics}).

\begin{table}[t]
\centering
\caption{Ranking of designs under the proxy and the ordinal model.}
\label{tab:proxy-ranking}
\small
\begin{tabular}{@{}ll
  S[table-format=1.2]S[table-format=1.2]
  S[table-format=2.0]S[table-format=2.0]
  S[table-format=1.1]S[table-format=2.1]@{}}
\toprule
 & & \multicolumn{2}{c}{Rank correlation}
   & \multicolumn{2}{c}{Same best design (\%)}
   & \multicolumn{2}{c}{Regret (\%)} \\
\cmidrule(lr){3-4}\cmidrule(lr){5-6}\cmidrule(l){7-8}
 & & {Proxy} & {Benchmark} & {Proxy} & {Benchmark} & {Proxy} & {Benchmark} \\
\midrule
\multicolumn{8}{@{}l}{\emph{A. Allocations, by scale and responses per item}} \\
Binary     & 10  & 0.89 & 0.91 & 62 & 60 & 1.5 & 0.8 \\
           & 30  & 0.91 & 0.91 & 58 & 45 & 0.5 & 0.7 \\
           & 100 & 0.92 & 0.93 & 47 & 52 & 0.3 & 0.2 \\
           & 300 & 0.91 & 0.91 & 45 & 38 & 0.1 & 0.1 \\
\addlinespace[3pt]
Five-point & 10  & 0.93 & 0.92 & 60 & 52 & 0.7 & 0.7 \\
           & 30  & 0.90 & 0.90 & 37 & 45 & 0.6 & 0.5 \\
           & 100 & 0.91 & 0.92 & 35 & 45 & 0.1 & 0.1 \\
           & 300 & 0.89 & 0.89 & 30 & 27 & 0.1 & 0.1 \\
\midrule
\multicolumn{8}{@{}l}{\emph{B. Formats, by completion profile}} \\
Mild       &     & 0.92 & 0.92 & 95 & 95 & 0.5 & 0.6 \\
Moderate   &     & 0.79 & 0.80 & 55 & 53 & 4.3 & 3.9 \\
Steep      &     & 0.74 & 0.72 & 63 & 61 & 6.1 & 6.3 \\
\bottomrule
\end{tabular}
\end{table}

\subsection{Planning the questionnaire}
\label{subsec:num-allocation}
\begin{figure}[t]
\centering
\includegraphics{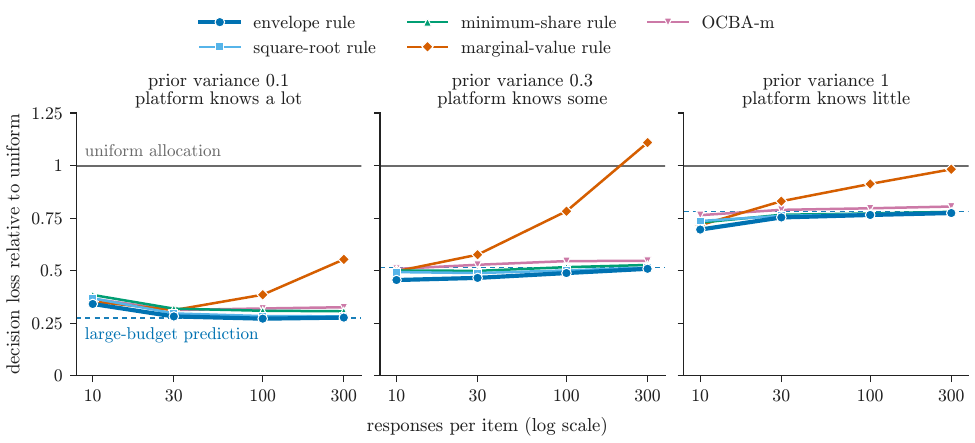}
\caption{Decision loss of each planning rule relative to uniform
allocation, averaged over the three prior configurations and the two
scales. The dashed line is the large-budget prediction computed from the
prior alone.}
\label{fig:num-allocation}
\end{figure}

The envelope rule does best among the rules we consider, and its gain
over uniform allocation grows as the prior variance falls
(Figure~\ref{fig:num-allocation}). We compare it with five other
planning rules.
\begin{enumerate}
\item \emph{Uniform allocation.} Every item receives the same number of
responses.
\item \emph{Square-root rule} (Appendix~\ref{app:proofs-alloc}). Responses
follow \eqref{eq:sqrt-rule}, the large-budget plan of
Proposition~\ref{prop:sqrt-law}.
\item \emph{Marginal-value rule} (Section~\ref{subsec:envelope-rule}).
Each response goes where it lowers the proxy loss most.
\item \emph{Minimum-share rule.} Every item first receives one fifth of
its even share, and the marginal-value rule allocates the rest.
\item \emph{OCBA-m} \citep{ChenHeFuLee2008}. This standard rule for
selecting a top subset in simulation sets each item's total information,
from the prior and its responses, inversely proportional to the expected
squared distance of its value from the boundary between the \(K\)th
and \((K+1)\)th prior means.
\end{enumerate}

Figure~\ref{fig:num-allocation} shows the decision loss of each rule
relative to uniform allocation, with 40 prior instances per
configuration and scale and 200 replications per instance
(standard errors in Appendix~\ref{app:numerics}). The envelope rule has
the lowest or tied-lowest loss at every prior variance and budget, and
no rule does significantly better in any of the 72 settings of
configuration, scale, prior variance, and budget. It lowers
the decision loss by 56 to 81 percent with prior variance \(0.1\), by
about half with \(0.3\), and by 15 to 43 percent with \(1\). The
square-root rule, which uses only large-budget cutoff weights, attains 86 to 100 percent of the envelope rule's improvement. Most of
the gain thus comes from favoring items near the cutoff. The envelope
rule's advantage over the square-root rule is up to 4 percent of the
uniform loss with 10 responses per item and not significant with 300
(Appendix~\ref{app:numerics}).

The myopic marginal-value rule does well with small budgets but falls
behind as the budget grows. With prior variance
\(0.3\) and 300 responses per item, its loss exceeds that of uniform
allocation. The minimum-share rule spends part of the budget on items whose
place is already clear from the prior. OCBA-m
does better than uniform allocation in every setting and attains 70 to
99 percent of the envelope rule's improvement. Its shares fall only with
the squared distance from the boundary, so it spends more of the budget
than the envelope rule on items far from the cutoff.

To see how close the envelope rule comes to the best plan, we repeat
the comparison with \(L=12\) items and \(K=3\) in the random
configuration and search for the best fixed allocation under the
proxy by numerical optimization from many starting points. The envelope rule
attains on average 100 to 105 percent of the best allocation's
improvement over uniform allocation, and at least 94 percent in every
setting. It can exceed 100 percent because the best allocation is
found under the proxy.

Proposition~\ref{prop:sqrt-law} predicts the envelope rule's
large-budget gains well. The dashed lines in
Figure~\ref{fig:num-allocation} show the predicted loss ratio,
computed from the prior alone and averaged over the three
configurations. With 300 responses per item, the envelope rule's loss ratio is
within 0.03 of this prediction in all nine combinations of
configuration and prior variance.

The results change little with the problem size or the rule's
numerical settings. With \(L=30\) items and \(K=3\), and with \(L=120\)
items and \(K=6\) or \(12\) (random configuration, prior variance \(0.3\)
and \(1\), binary ratings), the envelope rule has the lowest loss in
23 of the 24 settings and is never significantly worse than another rule. Refreshing the cutoffs after every 5 or 25 percent growth
instead of 10, or simulating them with 1{,}000 or 4{,}000 draws instead of
2{,}000, changes its loss by at most 1.2 percent of the uniform loss.
Planning all four budgets takes about 4, 10, and 20 seconds on a laptop
with 30, 60, and 120 items.

\subsection{Which items each plan asks about}
\label{subsec:num-exploration}
\begin{figure}[t]
\centering
\includegraphics{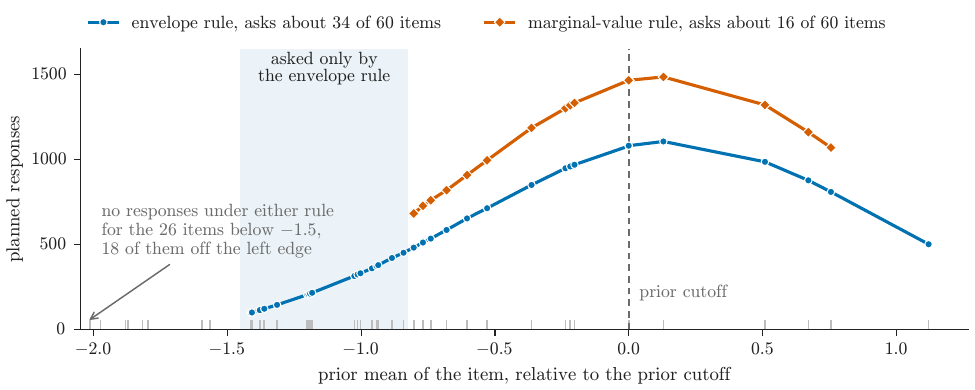}
\caption{Planned responses per item in one instance, by the item's prior
mean relative to the prior cutoff. Gray ticks mark the candidate
items.}
\label{fig:allocation-by-mean}
\end{figure}

Figure~\ref{fig:allocation-by-mean} shows why the marginal-value rule
falls behind, using one instance of the random configuration with prior
variance \(0.3\), binary ratings, and 300 responses per item. The marginal-value rule asks about
only the 16 of the 60 items closest to the prior cutoff. The envelope rule asks about 34 items and gives smaller blocks
to those well below the cutoff. The marginal-value rule never asks
about these items (shaded band), although they could be selected. Under the proxy, the loss is 0.068 for the
marginal-value plan, 0.060 for uniform allocation, and 0.026 for the
envelope plan.

Across instances with prior variance \(0.3\) and 300
responses per item, the marginal-value rule leaves on average 35 of the
60 items without a response, and the envelope rule leaves 15. These
15 are items whose place is already clear from the
prior. With prior variance \(0.1\), the envelope rule asks about only 9 to 23
items, depending on the budget.

\subsection{Choosing the format}
\label{subsec:num-format}\label{subsec:eis-screening}
Choosing the format by EIS comes within 2 percent of the best format.
Rules of thumb have 14 to 88 percent higher loss
(Table~\ref{tab:num-format}). We test this under the ordinal model with
the five formats of Table~\ref{tab:eis-calibrated} and five completion
profiles. Three profiles are synthetic, with mild, moderate, and steep completion
loss. Across them, 80 to 90 percent of customers complete the lightest
format (three binary questions) and 69, 36, or 17 percent the
heaviest (seven five-point questions). The other two are the
calibrated scenarios of Section~\ref{subsec:eis-calibration}. For each
format, we show the questionnaire to 3, 10, or 30 customers per item
and plan the allocation with the envelope rule.

The relative information \(\beta(S)\) depends on where the cutoff lies
(Section~\ref{subsec:eis-calibration}), so we compute it from
each instance's prior. In the random configuration, where the cutoff is far
above the threshold of the binary scale, a five-point rating provides
2.2 to 3.2 times the information of a binary one at the cutoff. The
ratio is about 1.5 in the separated and crowded configurations, whose
cutoffs are closer to that threshold.

We compare four rules that choose the format with the largest EIS, the
most responses, the finest scale with the longest questionnaire, or the
lightest questionnaire. Table~\ref{tab:num-format} reports the decision
loss of each relative to the best format in hindsight, averaged
over configurations, completion profiles, and numbers of customers, with
300 instances per setting. The format with the largest EIS is the best
in 67 to 82 percent of settings.

\begin{table}[t]
\centering
\caption{Ordinal decision loss of the chosen format relative to the best
format in hindsight. In the column marked ``with limit,'' no item
receives more responses than there are customers who complete the
questionnaire. Bold marks the lowest loss in each column.}
\label{tab:num-format}
\small
\begin{tabular}{@{}l*{4}{S[table-format=1.2]}@{}}
\toprule
 & \multicolumn{4}{c}{Prior variance} \\
\cmidrule(l){2-5}
Format chosen by & {0.1} & {0.1 with limit} & {0.3} & {1} \\
\midrule
Largest EIS                          & \bfseries 1.02 & \bfseries 1.03 & \bfseries 1.01 & \bfseries 1.00 \\
Most responses                       & 1.16 & 1.16 & 1.14 & 1.28 \\
Finest scale, longest questionnaire  & 1.15 & 1.22 & 1.16 & 1.14 \\
Lightest questionnaire               & 1.61 & 1.66 & 1.58 & 1.88 \\
\midrule
Largest EIS is the best format (\% of settings) & {67} & {69} & {71} & {82} \\
\bottomrule
\end{tabular}
\end{table}

The model drops the limit of one response per item from each customer who completes the questionnaire (Section~\ref{subsec:collection}). With prior variance \(0.1\), the envelope rule's plan concentrates on a few items and exceeds this limit in 25 percent of cases, and it almost never does so with \(0.3\) or \(1\). When the plan is recomputed under the limit, the format with the largest EIS stays within 3 percent of the best (Table~\ref{tab:num-format}), and within 4 percent where the limit binds.

\subsection{Relaxing the assumptions}
\label{subsec:num-robust}
The envelope rule keeps its advantage under the first three relaxations
below. We then measure the gain from sending the questionnaire in waves
and show how to guard against a wrong prior. Unless stated
otherwise, we use the random and crowded configurations with prior
variance \(0.3\) and \(1\).

\emph{Correlated items.} The prior correlation between items \(i\)
and \(j\), indexed in random order, is \(\rho^{|i-j|}\) with \(\rho=0.6\).
The rules plan as if the items were
independent. The platform computes its
estimates from the full correlated prior by expectation propagation
\citep{Minka2001}. The envelope rule lowers the decision loss
relative to uniform allocation by 42 to 58 percent with prior variance
\(0.3\) and by 15 to 35 percent with \(1\), much as it does with
independent items.

Planning with the full covariance does not help. In a
comparison with \(\rho=0.6\) and \(0.9\) (random configuration, prior
variance \(0.3\), binary ratings, 10 instances each), the envelope rule plans with the full covariance. This changes its loss relative to uniform allocation by \(-0.01\) to \(+0.13\). When every item can be asked about
directly, a response about a contested item is worth more than one about
a correlated item.

\emph{Uneven prior information.} Half of the twelve items nearest the prior
cutoff have one tenth of the usual prior variance, as
if the platform had already collected responses about them. With prior variance
\(0.3\) the envelope rule lowers the loss by 45 to 58 percent, and with
\(1\) by 13 to 41 percent. The marginal-value rule keeps asking
about the items the platform knows best and does up to 87 percent worse
than uniform allocation. A plan that brings every item to the same
residual variance asks most about the most uncertain items. Its loss is
0.94 to 1.16 times that of uniform allocation, because its responses go
to uncertain items far from the cutoff.

\emph{Informative missingness.} Customers with strong opinions are
more likely to answer. A customer with utility \(u_i\) answers with
probability \(0.2+0.6/(1+e^{-2(|u_i|-1)})\), and the platform's
estimates ignore this. The decision loss rises in the random configuration and falls in the
crowded one, but the envelope rule's advantage over uniform allocation
is unchanged.

\emph{Sending the questionnaire in waves.}\label{subsec:num-waves}
Platforms that survey customers after each delivery can send the
questionnaire in waves and use early responses to plan later waves. We split the budget into one, two, or four equal
waves (random configuration, binary ratings, 30 and 100 responses per
item). The envelope rule of Section~\ref{sec:allocation} plans the first wave
from the prior and each later wave from the current posterior means and
variances. Two waves
lower the decision loss relative to a single wave by 38 to 54 percent,
and four waves by 49 to 69 percent. Most of the gain comes from the
second wave.

\emph{A wrong prior.} Widening the prior and sending the questionnaire in waves both reduce
the loss from a wrong prior. We draw the true
values from a distribution other than the prior used
to plan and estimate. The true values have twice or four
times the prior variance, the prior means of a random quarter of the
items are shuffled among those items, or three items below the prior cutoff have true values centered
three prior standard deviations above their prior means. With 10 responses per item, the envelope rule does better
than uniform allocation in all cases but one. With 300 responses per item and prior variance
\(0.3\), its loss is up to 7.7 times the uniform loss, because the envelope rule never asks
about items the wrong prior places far from the cutoff, and uniform
allocation does. When the prior is right, doubling
its variance in planning and estimation costs at most 8 percent of the
uniform loss. When it is wrong, the envelope rule with doubled variance beats uniform
allocation except for shuffled means with 30 or more responses per item.
A second wave planned from first-wave responses makes the envelope rule
better than uniform allocation in every case with prior variance
\(1\). It helps less with \(0.3\), since the plan depends more on the
wrong prior (Appendix~\ref{app:numerics}).

\subsection{MovieLens ratings}
\label{subsec:num-movielens}\label{subsec:numerics_movielens}
On real ratings, too, targeting the cutoff lowers the decision loss,
but asking about the most uncertain movies does not. We test the
rules on the MovieLens 100K ratings \citep{HarperKonstan2016}. We order the ratings by
time and use the first 20 percent as past data. The 200 movies with the
most past ratings are the candidate items, and the platform selects
\(K=10\). The platform's prior for a movie is normal, centered at
its past mean rating, with variance equal to the variance of a single
rating divided by the number of past ratings. From this prior, each
rule plans how many new ratings to request for each movie, up to the
number available.

The remaining ratings of each movie are split at random into two
halves. One half is the pool of new ratings, and the
mean of the other half is the movie's true value. Splitting at random rather than by time matters little, since a movie's mean rating differs across time periods about as much as sampling noise predicts (variance ratios 1.01 to 1.04). We average the results over 40 random splits. In
each split the rules plan anew. Then, in each of 25 replications, every
movie's pool is shuffled and each movie gets its planned number of
ratings. The platform selects the ten movies with the highest mean
of their past and new ratings.

Table~\ref{tab:num-movielens} reports the decision loss relative to
uniform allocation for budgets of 2 to 10 new ratings per movie. The
envelope, minimum-share, and marginal-value rules lower the loss by 6 to
28 percent. Asking about the most uncertain movies does no better than
uniform allocation. The envelope rule does best with 5 and 8
ratings per movie, and the minimum-share rule with 2 and 10. At every
budget the envelope rule outperforms the square-root rule by 4 to 8
percent of the uniform loss. The standard errors of the paired
differences between rules are at most 2.3 percent of the uniform loss.
A single split by time, with the next 40 percent of the ratings as the
pool and the last 40 percent for the true values, gives similar gains
of 9 to 31 percent for the same three rules. Their ranking differs, as it
does between random splits.

\begin{table}[tb!]
\centering
\caption{Decision loss relative to uniform allocation on MovieLens
ratings, averaged over 40 random splits. Bold marks the lowest loss in
each column.}
\label{tab:num-movielens}
\small
\begin{tabular}{@{}l*{4}{S[table-format=1.2]}@{}}
\toprule
 & \multicolumn{4}{c}{New ratings per movie} \\
\cmidrule(l){2-5}
 & {2} & {5} & {8} & {10} \\
\midrule
Envelope rule         & 0.94 & \bfseries 0.82 & \bfseries 0.78 & 0.75 \\
Square-root rule      & 1.02 & 0.90 & 0.82 & 0.79 \\
Minimum-share rule    & \bfseries 0.91 & 0.87 & 0.85 & \bfseries 0.72 \\
Marginal-value rule   & 0.92 & 0.86 & 0.84 & 0.76 \\
Most uncertain movies & 0.99 & 1.03 & 1.06 & 1.06 \\
\bottomrule
\end{tabular}
\end{table}

\FloatBarrier
\section{Conclusion}
\label{sec:conclusion}

A platform that collects customer feedback before it selects which
candidate items to offer must decide which items to ask about and in
what format. Both decisions depend on the selection cutoff.
Responses matter most for uncertain items likely to be near the cutoff.
Allocating responses by uncertainty alone is no better than spreading
them evenly. For an item far from the
cutoff, the value of responses grows slowly over the first few and
faster after that. A plan should therefore consider blocks of responses,
as the envelope rule does. The gain of planning over an even spread can be predicted
from the prior before the questionnaire is sent. It is largest when the
prior is right and informative enough that few items are contested. But a plan built on an overconfident prior never asks
about items the prior wrongly places far from the cutoff. A platform
unsure of its prior should widen it and
send the questionnaire in waves. With
a correct prior, two waves also lower the decision loss by 38 to 54
percent relative to one wave.

Formats can be compared by their effective information supply, the
expected information at the cutoff per customer asked. A finer scale is better only if its extra information outweighs
the drop in completion.

The completion rates in our comparison of formats
are assumed, and a randomized experiment across formats would measure
them. The information a rating provides also
depends on where the scale's thresholds lie relative to the cutoff, and
wording questions to move them closer is left for future work. Our analysis
could be extended to personalized selections, where the questions worth
asking depend on each customer's cutoff.

\paragraph{Code availability.} The code that produces every numerical
result in this paper is available from the authors on request.

\appendix
\section{Details and Proofs for Section 3}\label{app:proofs-proxy}
\paragraph{Conditions of Proposition~\ref{prop:kappa-star}.}
Values are scaled so that the deviations have variance one. Write
\(\tau_i\) for the prior precision of item \(i\), one over its prior
variance, and let \(k:=\min(K,L-K)+1\). The proposition assumes two
conditions. (A1) The prior is independent across items, so that
\(\psi_i=f_i(\zeta_i)\). (A2) The prior precisions of the \(k\) least
precise items sum to more than one. Condition (A2) holds when every
prior variance is below \(k\), and the prior variances in this paper are
at most one. It keeps the ordinal limit in part (i) finite, as
part (ii) requires. The inverse information \(1/I_S(c)\) grows roughly
like \(e^{c^2/2}\) as \(c\) moves away from the thresholds. For the
cutoff that item \(i\) faces to lie far above the thresholds, at least
\(K\) other items must have values that far out, and for it to lie far
below, at least \(L-K\) of them must. Under (A2), the probability of
these events, multiplied by the density \(\psi_i\) of the item's own
value at the cutoff, falls faster than \(1/I_S\) grows. Under a
correlated prior, the value of an item and its cutoff move together, so
\(\psi_i\) need not be small when the cutoff is far out, and a bound on
the prior variances no longer ensures a finite limit.

\paragraph{Notation for the proof.}
Fix the scale \(S\). For an item whose value is \(v\), write
\(p_s(v):=\Phi(\theta_s-v)-\Phi(\theta_{s-1}-v)\) for the probability that a
response falls in category \(s\), as in~\eqref{eq:ordered-probit}, and
\(\mathbf p(v)\) for the vector of these probabilities. Write \(\hat w_i\)
for the posterior mean of \(w_i\) after \(n\) responses about every item.
Under~(A1) it depends only on the responses about item \(i\). For the
proxy, replace these \(n\) responses by \(n\) independent normal signals
with mean \(w_i\) and variance \(1/\kappa_i\), and let \(\tilde w_i\) be the
posterior mean of \(w_i\) given them. The vector \(\tilde\w\) then has the
law~\eqref{eq:proxy-postmean}, so this construction leaves
\(\Delta(\N;S)\) unchanged. Both \(\hat w_i\) and \(\tilde w_i\) depend on
\(n\).

For a number \(c\), thought of as a cutoff, a near miss on item \(i\) occurs when the estimate and the value of the item lie on
different sides of \(c\), and it costs the distance
\(|w_i-c|\). The near-miss cost of item \(i\) at \(c\) is the expected cost,
\begin{equation}\label{eq:p1-near-miss}
g_{i,n}(c):=\E\big[\,|w_i-c|\;\1\{\hat w_i\text{ and }w_i\text{ lie on
different sides of }c\}\big],
\end{equation}
where an estimate equal to \(c\) counts as above \(c\). Define
\(\tilde g_{i,n}(c)\) likewise, with \(\tilde w_i\) in place of \(\hat w_i\).

\begin{lemma}[Loss decomposition by item]\label{lem:p1-split}
Suppose~(A1). For each item \(i\), let \(\mathbf y^{(i)}\) be the vector
that holds the estimates \(\hat w_1,\ldots,\hat w_i\) of the first \(i\)
items and the values \(w_{i+1},\ldots,w_L\) of the others, and let
\(\hat\zeta_i:=\zeta_i(\mathbf y^{(i)})\) be the cutoff that item \(i\) faces
in this vector. Then \(\hat\zeta_i\) is independent of \((w_i,\hat w_i)\),
and \(\Delta_{\mathrm{ord}}(\N;S)=\sum_{i=1}^L\E[g_{i,n}(\hat\zeta_i)]\) for
every \(n\). The same holds for \(\Delta(\N;S)\), with \(\tilde\w\) in place
of \(\hat\w\), the cutoff \(\tilde\zeta_i\), and
\(\tilde g_{i,n}\) in place of \(g_{i,n}\).
\end{lemma}

\paragraph{Proof.}
For any vector \(\mathbf x\) and item \(i\),
\begin{equation}\label{eq:p1-VK-split}
V_K(\mathbf x)=V_K(\mathbf x_{-i})+\big(x_i-\zeta_i(\mathbf x)\big)^+,
\end{equation}
where \(V_K(\mathbf x_{-i})\) is the total of the \(K\) largest entries of
\(\mathbf x\) without item \(i\). Indeed, if \(x_i\ge\zeta_i(\mathbf x)\),
the \(K\) largest entries of \(\mathbf x\) are \(x_i\) and the \(K-1\)
largest entries without item \(i\), which total
\(V_K(\mathbf x_{-i})-\zeta_i(\mathbf x)\), and otherwise they are the \(K\)
largest entries without item \(i\). Ties do not affect the identity.

Let \(\mathbf y^{(0)}:=\w\), so that \(\mathbf y^{(L)}=\hat\w\). The vectors
\(\mathbf y^{(i-1)}\) and \(\mathbf y^{(i)}\) differ only in entry \(i\),
which is \(w_i\) in the first and \(\hat w_i\) in the second, so they share
the cutoff \(\hat\zeta_i\) and~\eqref{eq:p1-VK-split} gives
\(V_K(\mathbf y^{(i-1)})-V_K(\mathbf y^{(i)})=(w_i-\hat\zeta_i)^+-(\hat w_i-\hat\zeta_i)^+\).
Summing over \(i\) gives
\(V_K(\w)-V_K(\hat\w)=\sum_{i=1}^L\big[(w_i-\hat\zeta_i)^+-(\hat w_i-\hat\zeta_i)^+\big]\)
for every realization. Under~(A1) the pairs \((w_j,\hat w_j)\),
\(j=1,\ldots,L\), are independent, because \(\hat w_j\) is a function of
the responses about item \(j\), which depend only on \(w_j\) and on the
deviations of the customers asked about it. The cutoff \(\hat\zeta_i\) depends only on the
pairs of the other items, so it is independent of \((w_i,\hat w_i)\). Each
term is at most \(|w_i-\hat w_i|\) in absolute value, which is integrable,
so taking expectations and conditioning on \(\hat\zeta_i\) gives
\(\Delta_{\mathrm{ord}}(\N;S)=\sum_i\E[d_i(\hat\zeta_i)]\) with
\(d_i(c):=\E[(w_i-c)^+]-\E[(\hat w_i-c)^+]\).

It remains to show that \(d_i=g_{i,n}\). Given the responses, the value
\(w_i\) has mean \(\hat w_i\). If \(\hat w_i\ge c\), then
\((\hat w_i-c)^+\) is the conditional mean of \(w_i-c\), so the conditional
mean of \((w_i-c)^+\) exceeds \((\hat w_i-c)^+\) by the conditional mean of
\((c-w_i)^+\), the expected distance by which the value falls below \(c\).
If \(\hat w_i<c\), the excess is the conditional mean of \((w_i-c)^+\), the
expected distance by which the value lies above \(c\). In both cases the
excess is the conditional expected near-miss cost, and averaging over the
responses gives \(d_i(c)=g_{i,n}(c)\). The argument for the proxy is the
same.
\hfill\(\square\)

\begin{lemma}[Monotonicity of the estimate]\label{lem:p1-monotone}
Fix an item and omit its index. The posterior mean \(\hat w\) is
nondecreasing in each response. Let response \(j\) be the category into
which \(w+\varepsilon_j\) falls, where \(\varepsilon_j\) is the
deviation~\eqref{eq:utility} of the \(j\)th customer asked about the item.
For fixed \(\varepsilon_1,\ldots,\varepsilon_n\), \(\hat w\) is then
nondecreasing in \(w\). Hence, for every \(c\), the probability that
\(\hat w<c\) is nonincreasing in \(w\), and that of \(\hat w\ge c\) is
nondecreasing.
\end{lemma}

\paragraph{Proof.}
Let \(Z\) be standard normal. Because \(\int_a^bz\phi(z)\,dz=\phi(a)-\phi(b)\),
the derivative of \(\log p_s(v)\) in \(v\) is the mean of \(Z\) given
\(\theta_{s-1}-v<Z\le\theta_s-v\). This truncated mean rises when either
end of the interval rises, so for \(s'>s\) the ratio \(p_{s'}(v)/p_s(v)\)
is nondecreasing in \(v\). Raising one response from \(s\) to \(s'\)
multiplies the posterior density by this ratio, up to a constant, so the
new posterior dominates the old one in the likelihood-ratio order and its
mean is at least as large. A larger \(w\) raises every
\(w+\varepsilon_j\) and so weakly raises every response, which gives the
remaining claims. \hfill\(\square\)

\begin{lemma}[Uniform local normality]\label{lem:p1-local}
Fix an item, omit its index, write \(\Pr_v\) for probabilities when its
value is \(v\), and fix a closed bounded interval \(W\).
\begin{enumerate}[label=(\alph*)]
\item For every sequence \(v_n\) in \(W\) that converges to some
\(v_0\in W\), the distribution of
\(\sqrt n(\hat w-v_n)\) under \(\Pr_{v_n}\) converges to
\(\mathcal N\big(0,1/I_S(v_0)\big)\).
\item There are constants \(\delta,a,C>0\) and \(n_0\) such that
\(\Pr_v\big(\sqrt n\,|\hat w-v|\ge t\big)\le Ce^{-at^2}\) for all
\(n\ge n_0\), \(v\in W\), and \(0\le t\le\delta\sqrt n\).
\end{enumerate}
\end{lemma}

\paragraph{Proof.}
Let \(\hat{\mathbf q}=(\hat q_1,\ldots,\hat q_S)\) be the fractions of the
\(n\) responses that fall in each category, and write \(\|\cdot\|\) for the
largest absolute entry of a vector. The posterior depends on the
responses only through \(\hat{\mathbf q}\). For a vector of fractions
\(\mathbf q\), the log-likelihood per response is
\(\ell_{\mathbf q}(v):=\sum_sq_s\log p_s(v)\), and the posterior density of
the value is proportional to \(e^{-G(v)}\) with
\(G(v):=\tau(v-\mu)^2/2-n\,\ell_{\hat{\mathbf q}}(v)\).

\emph{Maximum likelihood.} Each \(\log p_s\) is concave, because its
derivative, the truncated normal mean in the proof of
Lemma~\ref{lem:p1-monotone}, falls as \(v\) rises. Hence
\(\ell_{\mathbf q}\) is concave and \(G\) is convex. At
\(\mathbf q=\mathbf p(v)\), \(\ell_{\mathbf q}'(v)=\sum_sp_s'(v)=0\) and
\(-\ell_{\mathbf q}''(v)=I_S(v)>0\). Let \(I_{\min}>0\) be the smallest value of \(I_S\) on \(W\). By the implicit function theorem and
the compactness of \(W\), there are \(\eta,\delta_1>0\), with \(\eta\) below
half the smallest category probability \(p_s(v)\) over \(v\in W\), and a
continuously differentiable map \(m\) such that, whenever
\(\|\mathbf q-\mathbf p(v)\|\le\eta\) for some \(v\in W\), the function
\(\ell_{\mathbf q}\) has a unique maximizer \(m(\mathbf q)\), with
\(m(\mathbf p(v))=v\), and \(-\ell_{\mathbf q}''\ge I_{\min}/2\) within
\(2\delta_1\) of \(m(\mathbf q)\). At \(\mathbf p(v)\), the derivative of \(m\) in
\(q_s\) is \((\log p_s)'(v)/I_S(v)\). Let \(\Lambda\) be the largest change in
\(m\) per unit change of \(\mathbf q\) on this set.

\emph{Posterior mean and maximum likelihood.} There are \(C_1\) and
\(n_0\) such that \(|\hat w-m(\hat{\mathbf q})|\le C_1/n\) whenever
\(\|\hat{\mathbf q}-\mathbf p(v)\|\le\eta\) for some \(v\in W\) and \(n\ge n_0\).
This is Laplace's approximation of a posterior mean
\citep{tierney1986accurate,KassTierneyKadane1990}, and its error is uniform over this set.
Within \(2\delta_1\) of \(m(\hat{\mathbf q})\), the convex function \(G\) has
second derivative between \(nI_{\min}/2\) and \(n\Gamma\) and third
derivative at most \(n\Gamma\) in absolute value, where \(\Gamma\) depends
only on \(W\) and \(\eta\), and the prior term moves the minimizer of \(G\)
by \(O(1/n)\). The expansion of the posterior mean around this
minimizer therefore has an error of at most \(C_1/n\), with constants that
depend only on these bounds.

\emph{Part (a).} Under \(\Pr_{v_n}\), \(\hat{\mathbf q}\to\mathbf p(v_0)\) in
probability, and \(\sqrt n\big(\hat{\mathbf q}-\mathbf p(v_n)\big)\)
converges in distribution to a normal vector with mean zero and
covariance \(\diag(\mathbf p(v_0))-\mathbf p(v_0)\mathbf p(v_0)^\top\) by
the Lindeberg--Feller central limit theorem
\citep[Proposition~2.27]{vanderVaart1998}. By the previous step and the
mean value theorem,
\(\sqrt n(\hat w-v_n)=\sqrt n\big(m(\hat{\mathbf q})-m(\mathbf p(v_n))\big)+o_p(1)\)
converges to a normal variable with mean zero and variance
\(\sum_sp_s\big((\log p_s)'/I_S\big)^2=1/I_S\) at \(v_0\), where we used
\(\sum_sp_s(\log p_s)'=0\).

\emph{Part (b).} By Hoeffding's inequality,
\(\Pr_v\big(|\hat q_s-p_s(v)|\ge x\big)\le2e^{-2nx^2}\) for each category.
If \(\|\hat{\mathbf q}-\mathbf p(v)\|\le\eta\), then
\(|\hat w-v|\le\Lambda\|\hat{\mathbf q}-\mathbf p(v)\|+C_1/n\). For \(t\ge1\)
and \(n\ge4C_1^2\), the event \(\sqrt n\,|\hat w-v|\ge t\) therefore
requires \(\|\hat{\mathbf q}-\mathbf p(v)\|>\eta\) or
\(\sqrt n\,\|\hat{\mathbf q}-\mathbf p(v)\|\ge t/(2\Lambda)\), whose
probabilities are at most \(2Se^{-2n\eta^2}\) and \(2Se^{-t^2/(2\Lambda^2)}\).
For \(t\le\delta\sqrt n\) the first is at most \(2Se^{-2\eta^2t^2/\delta^2}\).
For \(t<1\) the bound holds once \(C\ge e^{a}\). \hfill\(\square\)

\begin{lemma}[Limit at a fixed cutoff]\label{lem:p1-fixed}
Fix an item \(i\), a number \(c\), and any sequence \(c_n\to c\). Then
\[
n\,g_{i,n}(c_n)\to\frac{f_i(c)}{2I_S(c)}
\qquad\text{and}\qquad
n\,\tilde g_{i,n}(c_n)\to\frac{f_i(c)}{2\kappa_i}.
\]
Moreover, \(n\,\tilde g_{i,n}(c)\) is bounded uniformly in \(c\) and \(n\).
\end{lemma}

\paragraph{Proof.}
Omit the index \(i\). For a value \(v\) and a cutoff \(c'\), let
\(\pi_n(v;c')\) be the probability, given the value \(v\), that \(\hat w\)
and \(v\) lie on different sides of \(c'\), so that
\(g_n(c')=\int|v-c'|\,\pi_n(v;c')\,f(v)\,dv\). Apply
Lemma~\ref{lem:p1-local} with \(W=[c-1,c+1]\), reduce its \(\delta\) to at
most \(1/3\), and take \(n\) large enough that \(|c_n-c|\le\delta\).

\emph{Values far from the cutoff.} If \(v>c_n+\delta\),
Lemma~\ref{lem:p1-monotone}, and then Lemma~\ref{lem:p1-local}(b) with
\(t=\delta\sqrt n\), give
\(\pi_n(v;c_n)=\Pr_v(\hat w<c_n)\le\Pr_{c_n+\delta}(\hat w<c_n)\le Ce^{-a\delta^2n}\).
The case \(v<c_n-\delta\) is symmetric, so these values contribute at
most \(nCe^{-a\delta^2n}\,\E|w-c_n|\to0\) to \(n\,g_n(c_n)\).

\emph{Values near the cutoff.} With \(v=c_n+u/\sqrt n\), the values
within \(\delta\) of \(c_n\) contribute
\[
\int_{|u|\le\delta\sqrt n}|u|\;\pi_n\big(c_n+u/\sqrt n;\,c_n\big)\,
f\big(c_n+u/\sqrt n\big)\,du .
\]
For fixed \(u>0\), a near miss is the event \(\sqrt n(\hat w-v)<-u\), whose
probability tends to \(\Phi\big(-u\sqrt{I_S(c)}\big)\) by
Lemma~\ref{lem:p1-local}(a), and the case \(u<0\) is symmetric. By
Lemma~\ref{lem:p1-local}(b), the integrand is at most
\(C|u|e^{-au^2}\max f\), which is integrable. Dominated convergence gives
the limit \(f(c)\int|u|\,\Phi\big(-|u|\sqrt{I_S(c)}\big)\,du=f(c)/(2I_S(c))\),
because \(\int_0^\infty x\,\Phi(-x)\,dx=1/4\).

\emph{Proxy.} Given the signals, \(w\) is normal with mean \(\tilde w\) and
variance \(\sigma_n^2:=1/(\tau+n\kappa)\), and \(\tilde w\) is normal with
mean \(\mu\) and variance \(1/\tau-\sigma_n^2\). Let \(h_n\) be the density of
\(\tilde w\), and let \(\chi(z):=\E[(z+Z)^+]-z^+\) for standard normal
\(Z\), a nonnegative function with \(\int\chi(z)\,dz=\E[Z^2]/2=1/2\). The
last step of the proof of Lemma~\ref{lem:p1-split} gives
\(\tilde g_n(c)=\E\big[\sigma_n\,\chi\big((\tilde w-c)/\sigma_n\big)\big]\),
so \(n\,\tilde g_n(c)=n\sigma_n^2\int\chi(z)\,h_n(c+\sigma_n z)\,dz\).
Because \(n\sigma_n^2\to1/\kappa\) and \(h_n\to f\) uniformly, this tends to
\(f(c)/(2\kappa)\) along \(c_n\to c\). It is at most \(n\sigma_n^2\max h_n/2\),
which is bounded for all \(n\ge1\) because \(n\sigma_n^2\le1/\kappa\) and the
variance of \(\tilde w\) is at least \(\kappa/(\tau(\tau+\kappa))\).
\hfill\(\square\)

\begin{lemma}[Uniform bound]\label{lem:p1-envelope}
Fix an item \(i\) and \(\gamma>0\). There is a constant \(C\) such that,
for all \(n\ge1\) and all \(c\),
\(n\,g_{i,n}(c)\le C\,e^{(1/2+\gamma)c^2}f_i(c)\).
\end{lemma}

\paragraph{Proof.}
Omit the index \(i\) and write \(\sigma^2:=1/\tau\).
Among all rules that decide from the responses whether to treat the value
as above or below \(c\), the rule that compares \(\hat w\) with \(c\) has
the smallest expected near-miss cost. Given the responses, treating the
value as above \(c\) costs the conditional mean of \((c-w)^+\), treating it
as below costs the conditional mean of \((w-c)^+\), and the second minus
the first is \(\hat w-c\). Hence \(g_n(c)\) is at most the expected
near-miss cost of any other rule.

\emph{Bounded cutoffs.} By Lemma~\ref{lem:p1-fixed}, \(n\,g_n\) converges to
the continuous function \(f/(2I_S)\) along every convergent sequence of
cutoffs, and hence uniformly on bounded intervals. With
\(g_n(c)\le\E|w-c|\) for small \(n\), this shows that \(n\,g_n\) is bounded
on any \([-c_0,c_0]\) uniformly in \(n\), and the claim holds there because
\(e^{(1/2+\gamma)c^2}f(c)\) is bounded away from zero on this interval.

\emph{High cutoffs.} Let \(\theta:=\theta_{S-1}\) be the highest threshold,
\(x:=c-\theta\), and \(r(v):=\Phi(\theta-v)\), the probability that a
response about an item with value \(v\) falls below the top category. We
use the normal tail bounds \(\phi(y)/(2y)\le\Phi(-y)\le\phi(y)/y\) for
\(y\ge1\), and take \(c_0\) large enough that \(c\ge c_0\) implies
\(c\ge\mu+1\), \(x\ge2\), and \(r(c-1)\ge2r(c)\). The rule that always
treats the value as below \(c\) costs
\(\E[(w-c)^+]=\sigma^2f(c)-(c-\mu)\Pr(w>c)\le\sigma^2f(c)\), which proves the
claim when \(n\le e^{(1/2+\gamma)c^2}\).

Another rule treats the value as above \(c\) when the number \(X\) of
responses below the top category is at most \(nr(c)\). Given the value
\(v\), \(X\) is binomial with \(n\) trials and success probability \(r(v)\).
For success probabilities \(p\) and \(p'\), let
\(\mathrm{KL}(p,p'):=p\log(p/p')+(1-p)\log\big((1-p)/(1-p')\big)\) be the
relative entropy between the two Bernoulli distributions, which satisfies
\(\mathrm{KL}(p,p')\ge(p-p')^2/(2\max\{p,p'\})\). By Hoeffding's inequality in
its relative-entropy form \citep{Hoeffding1963}, the rule errs with
probability at most \(\exp\big(-n\,\mathrm{KL}(r(c),r(v))\big)\). We bound
the expected near-miss cost over three ranges of values.
\begin{enumerate}[label=(\arabic*)]
\item For \(v\ge c\), the difference \(r(c)-r(v)\) is at least
\((v-c)\phi(x+1)\) when \(v\le c+1\) and at least \(\phi(x+1)\) beyond,
while \(r(c)\le\phi(x)\). With \(\beta:=\phi(x+1)^2/\phi(x)\), the exponent
is at least \(n\beta\min\{(v-c)^2,1\}/2\). Since \(f(v)\le f(c)\) here,
this range costs at most
\(f(c)/(n\beta)+e^{-n\beta/2}\sigma^2f(c)\le(1+\sigma^2)f(c)/(n\beta)\).
\item For \(c-1\le v<c\), we have \(r(v)-r(c)\ge(c-v)\phi(x)\),
\(r(v)\le\phi(x-1)\), and \(\phi(x)^2/\phi(x-1)\ge\beta\), so the exponent
is at least \(n\beta(c-v)^2/2\). Since
\(f(v)\le f(c-1)\le e^{\tau(c-\mu)}f(c)\), this range costs at most
\(e^{\tau(c-\mu)}f(c)/(n\beta)\).
\item For \(v<c-1\), we have \(r(v)\ge r(c-1)\ge2r(c)\), so
\(r(v)-r(c)\ge r(v)/2\) and the exponent is at least \(nr(c-1)/8\). This
range costs at most \(e^{-nr(c-1)/8}\,\E|w-c|\).
\end{enumerate}
Because \(1/\beta=\sqrt{2\pi}\,e^{x^2/2+2x+1}\), ranges~(1) and~(2)
contribute at most \(Ce^{(1/2+\gamma)c^2}f(c)\) to \(n\,g_n(c)\) for every
\(n\). For range~(3), suppose \(n>e^{(1/2+\gamma)c^2}\). Then
\(nr(c-1)/8\ge e^{(1/2+\gamma)c^2}\phi(x-1)/\big(16(x-1)\big)\ge e^{\gamma c^2/2}\)
once \(c_0\) is large, and
\(n\,e^{-nr(c-1)/8}\le\big(8/r(c-1)\big)e^{-nr(c-1)/16}\). The resulting
factor \(\exp\big(-e^{\gamma c^2/2}/2\big)\) falls faster than
\(\E|w-c|/\big(r(c-1)f(c)\big)\) grows, so after enlarging \(c_0\), range~(3)
adds at most \(f(c)\) to \(n\,g_n(c)\).

\emph{Low cutoffs.} The case \(c\le-c_0\) is the mirror image, with the
lowest threshold \(\theta_1\) and the probability that a response falls
above the bottom category. \hfill\(\square\)

\paragraph{Proof of Proposition~\ref{prop:kappa-star}.}
By Lemma~\ref{lem:p1-split}, \(n\,\Delta_{\mathrm{ord}}(\N;S)\) and
\(n\,\Delta(\N;S)\) are the sums over items of
\(\E[n\,g_{i,n}(\hat\zeta_i)]\) and \(\E[n\,\tilde g_{i,n}(\tilde\zeta_i)]\),
respectively. Fix an item \(i\).

\emph{Step 1, the cutoff converges.} The \(K\)th largest entry of a vector
changes by at most the largest change of any entry, so
\(|\hat\zeta_i-\zeta_i|\le\max_{j<i}|\hat w_j-w_j|\). Think of the \(n\)
responses about item \(j\) as the first \(n\) of an infinite sequence. By
the martingale convergence theorem, \(\hat w_j\) converges almost surely to
the mean of \(w_j\) given the whole sequence, which is \(w_j\) itself
because the fraction of responses in the lowest category converges to
\(\Phi(\theta_1-w_j)\), a strictly decreasing function of \(w_j\). Hence
\(\hat\zeta_i\to\zeta_i\) almost surely, and the same argument for the
normal signals gives \(\tilde\zeta_i\to\zeta_i\) almost surely.

\emph{Step 2, the limit for each realization.} The function \(g_{i,n}\)
does not depend on the other items, so in every realization in which
\(\hat\zeta_i\to\zeta_i\), Lemma~\ref{lem:p1-fixed} applies to the sequence
\(c_n:=\hat\zeta_i\), and likewise for the proxy. Hence, almost surely,
\(n\,g_{i,n}(\hat\zeta_i)\to f_i(\zeta_i)/(2I_S(\zeta_i))=\psi_i/(2I_S(\zeta_i))\)
and \(n\,\tilde g_{i,n}(\tilde\zeta_i)\to\psi_i/(2\kappa_i)\).

\emph{Step 3, the tails.} The limit can be taken inside the
expectation if the two sequences in Step~2 are uniformly integrable
\citep[Theorem~2.20]{vanderVaart1998}. The proxy sequence is bounded by
Lemma~\ref{lem:p1-fixed}. For the ordinal sequence, choose \(\gamma>0\) so
that the prior precisions of the \(k\) least precise items sum to at least
\(1+4\gamma\), which~(A2) allows. By Lemma~\ref{lem:p1-envelope},
\(0\le n\,g_{i,n}(\hat\zeta_i)\le H_i(\hat\zeta_i)\) with the continuous
function \(H_i(c):=Ce^{(1/2+\gamma)c^2}f_i(c)\). It therefore suffices that
\(\sup_n\E\big[H_i(\hat\zeta_i)\,\1\{|\hat\zeta_i|>M\}\big]\to0\) as
\(M\to\infty\).

Each entry of \(\mathbf y^{(i)}\) is a value \(w_j\) or a posterior mean
\(\hat w_j\), and both have normal tails with the prior variance. For the
posterior mean, Jensen's inequality for conditional expectations gives
\(\E\big[e^{\lambda(\hat w_j-\mu_j)}\big]\le\E\big[e^{\lambda(w_j-\mu_j)}\big]=e^{\lambda^2/(2\tau_j)}\)
for every \(\lambda\), so \(\Pr(\hat w_j-\mu_j\ge s)\le e^{-\tau_js^2/2}\)
for \(s\ge0\), and likewise for the lower tail. The cutoff \(\hat\zeta_i\)
exceeds \(t\) only if at least \(K\) of the other entries exceed \(t\), and
it falls below \(-t\) only if at least \(L-K\) of them fall below \(-t\).
These entries are independent, so for large \(t\) and every \(n\),
\(\Pr(\hat\zeta_i>t)\le\sum_J\prod_{j\in J}e^{-\tau_j(t-\mu_j)^2/2}\) and
\(\Pr(\hat\zeta_i<-t)\le\sum_{J'}\prod_{j\in J'}e^{-\tau_j(t+\mu_j)^2/2}\),
where \(J\) ranges over the sets of \(K\) items other than \(i\), and \(J'\)
over the sets of \(L-K\) such items. Each of the sums
\(\tau_i+\sum_{j\in J}\tau_j\) and \(\tau_i+\sum_{j\in J'}\tau_j\) covers at
least \(k\) items, so it is at least \(1+4\gamma\). Because
\(\log H_i(c)=(1/2+\gamma-\tau_i/2)c^2+O(|c|)\), it follows that
\(H_i(c)\Pr(|\hat\zeta_i|\ge t)\le C'e^{-\gamma t^2/2}\) for
\(t\le|c|\le t+1\) and \(t\ge t_0\), uniformly in \(n\). Summing over
\(t=M,M+1,\ldots\) gives
\(\E\big[H_i(\hat\zeta_i)\1\{|\hat\zeta_i|>M\}\big]\le C'\sum_{m\ge0}e^{-\gamma(M+m)^2/2}\),
which tends to zero.

\emph{Step 4, conclusion.} Steps~1 to~3 give
\(n\,\E[g_{i,n}(\hat\zeta_i)]\to\tfrac12\E[\psi_i/I_S(\zeta_i)]\) for the
ordinal model and \(n\,\E[\tilde g_{i,n}(\tilde\zeta_i)]\to\E[\psi_i]/(2\kappa_i)\)
for the proxy. The first
limit is finite, since \(n\,g_{i,n}\le H_i\) gives \(f_i/(2I_S)\le H_i\)
in the limit, and the tail bound of Step~3 also holds for \(\zeta_i\),
the cutoff with every estimate replaced by the value. Summing over items
proves part~(i).

\emph{Part (ii).} Both limits in part~(i) are positive and finite, so the
ratio \(\Delta(\N;S)/\Delta_{\mathrm{ord}}(\N;S)\) tends to one if and
only if the two limits are equal.
If \(\kappa_i=\kappa_0\) for every item, the second limit is
\(\tfrac12\E[\sum_i\psi_i]/\kappa_0\), which equals the first if and only if
\(\kappa_0=\E[\sum_i\psi_i]/\E[\sum_i\psi_i/I_S(\zeta_i)]=\kappa(S)\)
by~\eqref{eq:kappa}. Dividing the two limits in part~(i)
gives~\eqref{eq:kappa-ratio}. \hfill\(\square\)

\begin{corollary}[Allocations with fixed shares]\label{cor:p1-shares}
Suppose the conditions of Proposition~\ref{prop:kappa-star} hold, and let
the budget \(N_{\mathrm{tot}}\) grow while every item keeps a fixed
positive share of it, \(N_i/N_{\mathrm{tot}}\to p_i>0\). Write
\(c_i^{\mathrm{ord}}:=\tfrac12\E[\psi_i/I_S(\zeta_i)]\) for the ordinal limit
of item \(i\) in Step~4, and recall the limiting cutoff weight
\(c_i^\infty=\tfrac12\E[\psi_i]\) of Section~\ref{subsec:sqrt-law}.
Then \(N_{\mathrm{tot}}\,\Delta_{\mathrm{ord}}(\N;S)\to\sum_ic_i^{\mathrm{ord}}/p_i\)
and \(N_{\mathrm{tot}}\,\Delta(\N;S)\to\sum_ic_i^\infty/(p_i\kappa_i)\). With
the common value \(\kappa_i=\kappa(S)\), the ratio
\(\Delta(\N;S)/\Delta_{\mathrm{ord}}(\N;S)\) therefore tends to an average of
the item-level values
\(c_i^\infty/c_i^{\mathrm{ord}}=\E[\psi_i]/\E[\psi_i/I_S(\zeta_i)]\) divided by
\(\kappa(S)\), with weights proportional to \(c_i^{\mathrm{ord}}/p_i\), so the
limit lies between the smallest and the largest of these values divided
by \(\kappa(S)\).
\end{corollary}

\paragraph{Proof.}
Lemma~\ref{lem:p1-split} does not require equal numbers of responses, so
\(N_{\mathrm{tot}}\,\Delta_{\mathrm{ord}}(\N;S)\) is the sum over items of
\((N_{\mathrm{tot}}/N_i)\,\E[N_i\,g_{i,N_i}(\hat\zeta_i)]\), where \(g_{i,N_i}\) is the near-miss cost of item \(i\) with \(N_i\)
responses. Every \(N_i\) grows without bound, so Steps~1 and~2 of the
proof of Proposition~\ref{prop:kappa-star} apply with \(N_i\) in place of
\(n\) and give \(N_i\,g_{i,N_i}(\hat\zeta_i)\to\psi_i/(2I_S(\zeta_i))\) almost
surely. Step~3 applies unchanged, because
Lemma~\ref{lem:p1-envelope} holds for every number of responses and the
tail bound on \(\hat\zeta_i\) holds for every allocation. Since
\(N_{\mathrm{tot}}/N_i\to1/p_i\), the first limit follows, and the second
follows in the same way from Lemma~\ref{lem:p1-fixed}. Dividing the two
limits with \(\kappa_i=\kappa(S)\) gives the weighted average.
\hfill\(\square\)

\paragraph{Other budgets.}
At 30, 300, and 3,000 responses per item, which
Table~\ref{tab:proxy-loss} omits, the ratio of the proxy decision loss to
the ordinal decision loss is between 0.92 and 1.11 under the
information at the cutoff and between 0.41 and 1.57 under the per-item
prior average, with standard errors of at most 0.11 and 0.14.

\paragraph{Correlated items.}
We repeat the comparison of Section~\ref{subsec:proxy-validation} with
correlation \(\rho^{|i-j|}\) between items \(i\) and \(j\) and \(\rho=0.6\),
computing the ordinal posterior by expectation propagation
\citep{Minka2001}. From 10 to 1{,}000 responses per item and in every configuration and
scale, the ratio \(\Delta/\Delta_{\mathrm{ord}}\) is between \(0.89\) and
\(1.10\) under the information at the cutoff \eqref{eq:kappa} and
between \(0.44\) and \(1.60\) under the per-item prior average.

\paragraph{Item-level information at the cutoff.}
The common value \(\kappa(S)\) pools the information at the cutoff over
items. With the item-level value \(\E[\psi_i]/\E[\psi_i/I_S(\zeta_i)]\) in place
of \(\kappa(S)\), the two limits in Step~4 agree item by item. For the prior instances of
Section~\ref{sec:numerics}, the average difference between the two
values, weighted by the items' cutoff weights, is at most 3.3 percent
with binary ratings and 1.5 percent with five-point ratings. It is at
most 1.8 percent for the ratio of the five-point to the binary value,
which Section~\ref{sec:format} uses. The largest differences in a single
instance are 4.9, 2.3, and 2.6 percent. The two values are close
because every item faces about the same cutoff, the \(K\)th or
\((K+1)\)th largest value.

\section{Details and Proofs for Section 4}\label{app:proofs-alloc}

\paragraph{Information that differs across items.}
Propositions~\ref{prop:loss-representation}, \ref{prop:marginal-value},
\ref{prop:s-shape}, and~\ref{prop:envelope} hold with the same proofs
when the
information per response differs across items, with \(\kappa_i\) in place
of \(\kappa\). The loss is then
\(\sum_i(c_i/\kappa_i)/(\tau_i/\kappa_i+N_i)\). In
Proposition~\ref{prop:sqrt-law} and Corollary~\ref{cor:sqrt-rule},
\(c_i^\infty/\kappa_i\) therefore replaces \(c_i^\infty\) in the shares and
in \(L_{\mathrm{eff}}\), and the square-root rule \eqref{eq:sqrt-rule}
becomes \(\tau_i+\kappa_iN_i=\max\big(\tau_i,\sqrt{c_i^\infty\kappa_i}/\lambda\big)\).

\paragraph{The value curve under an independent prior.}
Under an independent prior, the estimate \(X_j(n)\) of
Section~\ref{subsec:exploration} is normal with mean \(\mu_j\) and
variance \(s_j(n)^2:=1/\tau_j-1/(\tau_j+\kappa n)\), and it is
independent of the cutoff \(\zeta_j(X)\). Hence
\begin{equation}\label{eq:item-value-closed}
\mathcal V_j(n)=\E\Big[s_j(n)\,\phi\Big(\frac{\mu_j-\zeta_j(X)}{s_j(n)}\Big)
+\big(\mu_j-\zeta_j(X)\big)\,\Phi\Big(\frac{\mu_j-\zeta_j(X)}{s_j(n)}\Big)\Big],
\end{equation}
with the expectation over the cutoff \(\zeta_j(X)\) only and the bracket
read as \((\mu_j-\zeta_j(X))^+\) when \(s_j(n)=0\). The rules evaluate this
expectation with simulated draws of the cutoffs.

\paragraph{Implementation of the rules.}
Both the marginal-value rule and the envelope rule hold the cutoffs
fixed within a batch of responses and recompute them each time the
planned total grows by the fraction \(g\). Each batch carries at least
the information \(\nu\) of \(L\) like/dislike responses
(Algorithm~\ref{alg:envelope}). Within a batch,
each item moves along its own value curve, so the gain of its next
response accounts for the responses already assigned to it. The curves
are evaluated at 0 to 24 additional responses and at 24 further points
spaced geometrically up to the batch size, and are interpolated linearly
in between. The envelope rule takes the concave envelope on these points
before interpolating. Ties are broken by a fixed random order of the
items. We use \(g=0.1\) and 2{,}000 draws of the cutoffs. The
experiments of Section~\ref{sec:numerics} set the batch minimum at \(L\)
responses on both scales rather than in information, which changes the
five-point losses by less than 0.5 percent.
Section~\ref{subsec:num-allocation} shows that \(g\) and the number of
draws matter little.

\paragraph{Proof of Lemma~\ref{lem:cutoff}.}
Applying the condition to the items in \(K\)th and \((K+1)\)th place
shows that neither value equals \(\bar w\), so
\(w_{(K)}>\bar w>w_{(K+1)}\). If item \(i\) is among the \(K\) items with
the highest values, then \(w_i\ge w_{(K)}>\bar w\) and
\(\hat w_i>w_i-(w_i-\bar w)=\bar w\). Otherwise \(w_i\le w_{(K+1)}<\bar w\)
and \(\hat w_i<w_i+(\bar w-w_i)=\bar w\). The \(K\) items with the highest
values therefore have estimates above \(\bar w\), and all other items have
estimates below it, so these \(K\) items have the highest
estimates. If every error is smaller than \((w_{(K)}-w_{(K+1)})/2\), the
condition holds, because \(|w_i-\bar w|\ge(w_{(K)}-w_{(K+1)})/2\) for every
\(i\). \hfill\(\square\)

\medskip
For estimates with covariance \(M\), the expected value of the selection
is \(G(M):=\E\big[V_K(\bmu+M^{1/2}Z)\big]\) with \(Z\) standard normal.

\begin{lemma}[Covariance path]\label{lem:covariance-path}
Fix positive semidefinite matrices \(M_0\) and \(D\) such that
\(M_0+tD\) is nonsingular for every \(t\in(0,1]\), and let
\(h(t):=G(M_0+tD)\). Then \(h\) is continuous and nondecreasing on
\([0,1]\) and continuously differentiable on \((0,1]\), with
\[
h'(t)=\tfrac12\trace\big(\E[\nabla^2V_K(X_t)]\,D\big),
\qquad X_t\sim\mathcal N(\bmu,M_0+tD),
\]
where the Hessian of the piecewise-linear function \(V_K\) is understood
as a measure, and
\[
G(M_0+D)-G(M_0)=\int_0^1h'(t)\,dt ,
\]
where the integral of the nonnegative \(h'\) is improper at \(t=0\)
when \(M_0\) is singular.
\end{lemma}

\paragraph{Proof.}
The function \(V_K\) is convex and Lipschitz, so \(h\) is finite, and it
is continuous because normal expectations of Lipschitz functions are
continuous in the covariance. For \(s<t\), \(X_t\) has the law of \(X_s\)
plus an independent normal shock with covariance \((t-s)D\), so
\(h(s)\le h(t)\) by Jensen's inequality. On \((0,1]\) the covariance is
nonsingular, so the normal density is smooth in the covariance. By the
Gaussian heat-equation identity, its derivative in the covariance is
half its Hessian in the argument, and integration by parts gives the
formula for \(h'(t)\). This derivative is continuous,
and nonnegative since \(h\) is nondecreasing. For every
\(t_0\in(0,1)\), the fundamental theorem of calculus gives
\(h(1)-h(t_0)=\int_{t_0}^1h'(t)\,dt\). As \(t_0\downarrow0\), the left side
converges to \(h(1)-h(0)\) by continuity, and the right side increases to
the integral. \hfill\(\square\)

\paragraph{Proof of Proposition~\ref{prop:loss-representation}.}
The true values \(\w\) and the estimates \(X=\tilde\w(\N)\) are normal
with covariances \(\bSigma\) and \(M(\N)=\bSigma-A(\N)\), so
\(\Delta(\N)=G(\bSigma)-G(M(\N))\). Along the path
\(M(\N)+tA(\N)=\bSigma-(1-t)A(\N)\), \(t\in[0,1]\), the covariance is at
least \(t\bSigma\) because \(A(\N)\le\bSigma\), so it is nonsingular for
\(t>0\). Lemma~\ref{lem:covariance-path} then gives
\(\Delta(\N)=\int_0^1\tfrac12\trace\big(\E[\nabla^2V_K(X_t)]A(\N)\big)\,dt\),
which is nonnegative because \(A(\N)\) is positive semidefinite. This
expression depends on the allocation only through \(A(\N)\).

Under an independent prior, \(A(\N)\) and every matrix on the path are
diagonal, so the integrand splits as
\(\sum_ia_i(\N)\,\tfrac12\E[\partial^2V_K(X_t)/\partial x_i^2]\). Fix
\(i\). By \eqref{eq:p1-VK-split}, \(V_K(X_t)=V_K(X_{t,-i})+(X_{t,i}-\zeta_i(X_t))^+\), so the second
derivative in \(x_i\) is a point mass where \(X_{t,i}\) crosses
\(\zeta_i(X_t)\), with expectation \(\E[f_{i,t}(\zeta_i(X_t))]\) because
\(X_{t,i}\) is independent of the other coordinates. Integrating
over \(t\) gives \(\Delta(\N)=\sum_ic_i(\N)a_i(\N)\). When \(N_i=0\), the
variance of \(X_{t,i}\) is \(t\,a_i(\N)\), so the integrand is of order
\(t^{-1/2}\) near \(t=0\) and the integral converges. \hfill\(\square\)

\paragraph{Proof of Proposition~\ref{prop:marginal-value}.}
Under an independent prior, the estimates \(X_i\) are independent,
\(X_j\) is normal with mean \(\mu_j\) and variance
\(s_j^2:=1/\tau_j-a_j(\N)\), and the law of the other estimates does not
depend on \(N_j\). By \eqref{eq:p1-VK-split},
\(V_K(X)=V_K(X_{-j})+(X_j-\zeta_j(X))^+\), so \(N_j\) affects \(\Delta(\N)\)
only through \(\E[(X_j-\zeta_j(X))^+]\), where \(X_j\) is independent of
the cutoff \(\zeta_j(X)\). For a fixed cutoff \(\zeta\) and \(Z\) standard
normal, the derivative of \(\E[(\mu_j+\sqrt v\,Z-\zeta)^+]\) in the
variance \(v>0\) is \(\phi\big((\mu_j-\zeta)/\sqrt v\big)/(2\sqrt v)\),
half the normal density at \(\zeta\). It is at most
\(1/(2\sqrt{2\pi v})\), so it can be taken inside the expectation over
the cutoff. Because \(ds_j^2/dN_j=\kappa a_j^2\), for \(N_j>0\) this gives
\(\mathrm{MV}_j(\N)=\kappa a_j^2\,b_j(\N)\) with
\begin{equation}\label{eq:local-weight}
b_j(\N)=\frac12\,\E\bigg[\frac{1}{s_j}\,
\phi\Big(\frac{\mu_j-\zeta_j(X)}{s_j}\Big)\bigg],
\end{equation}
half the density of \(X_j-\zeta_j(X)\) at zero. The argument holds
whatever the law of the cutoff, which has point masses at the prior
means of the other items without responses.

When \(N_j=0\), \(X_j=\mu_j\). Let \(q_j\) be the density of the law of
\(\zeta_j(X)\) after its point masses are removed, with \(q_j=0\) when no
other item has responses. It is bounded since it is at most the sum of the
densities of the estimates of other items with responses. It jumps
only at the prior means of the other items without responses, so it is
continuous at \(\mu_j\) unless one of them has prior mean \(\mu_j\), a tie
we rule out. The point masses then lie at a positive
distance from \(\mu_j\), where the kernel in \eqref{eq:local-weight} tends
to zero uniformly as \(N_j\downarrow0\), and the integral of the kernel
against \(q_j\) tends to \(q_j(\mu_j)\). Hence
\(b_j(\N)\to\tfrac12q_j(\mu_j)\), which is \(b_j\) at \(N_j=0\), half the
density of \(X_j-\zeta_j(X)\) at zero. Because \(\E[(X_j-\zeta_j(X))^+]\)
is continuous in \(N_j\), by the mean value theorem the right
derivative at \(N_j=0\) is the limit of \(\kappa a_j^2b_j(\N)\), which is
\(\kappa a_j^2b_j\) at \(N_j=0\). At \(\N=0\), the cutoff is the fixed
number \(\zeta_j(\bmu)\), so \(q_j=0\) and the derivative is zero for
every item.

By \eqref{eq:local-weight}, \(b_j=\tfrac12\E[f_j(\zeta_j(X))]\) for
\(N_j>0\), with \(f_j\) the density of \(X_j\). This is the integrand of
\(c_j(\N)\) at the start of the path in
Proposition~\ref{prop:loss-representation}. One more response moves
every point of the path except its end, but the loss depends only on the
two ends, so the marginal value involves only \(b_j\), the integrand at the start. As
\(\min_iN_i\to\infty\), \(X\) converges in law to \(\w\) and the densities
stay bounded, so \(b_j(\N)\to\tfrac12\E[\psi_j]=c_j^\infty\), the limit of
\(c_j(\N)\) in the proof of Proposition~\ref{prop:sqrt-law}.
\hfill\(\square\)

\paragraph{Properties of the local cutoff weight.}
Under an independent prior, \(b_j=\tfrac12q_j(\mu_j)\) at \(N_j=0\), which
is small when the prior mean lies far from the likely values of the cutoff.

The three properties in Section~\ref{subsec:exploration} hold exactly
for a cutoff fixed at \(\zeta\). Write
\(z_j:=\sqrt{\tau_j}\,|\mu_j-\zeta|\) and \(u:=\kappa N_j/\tau_j\), the
information in the responses relative to the prior. Then
\(s_j^2=u/\big(\tau_j(1+u)\big)\) and
\((\mu_j-\zeta)^2/s_j^2=z_j^2(1+u)/u\), so
\[
\log b_j=-\frac{z_j^2(1+u)}{2u}-\tfrac12\log u+\tfrac12\log(1+u)+\text{const},
\qquad
\frac{d\log b_j}{du}=\frac{z_j^2}{2u^2}-\frac{1}{2u(1+u)} .
\]
(i) If \(z_j>0\), then \(b_j\to0\) as \(u\to0\), because the first term,
of order \(-1/u\), dominates. (ii) If \(z_j>0\), the percentage growth rate of \(b_j\)
tends to infinity as \(u\to0\), whereas the second factor of
\eqref{eq:priority-decomposition}, proportional to
\((1+u)^{-2}\), falls at the percentage rate \(2/(1+u)\le2\). (iii) As
\(N_j\to\infty\), \(s_j^2\to1/\tau_j\) and \(b_j\) tends to half the prior
density of \(w_j\) at \(\zeta\). If \(z_j\ge1\), the derivative above is at
least \(1/\big(2u^2(1+u)\big)>0\), so \(b_j\) rises to this limit. If
\(0<z_j<1\), the derivative is positive for \(u<z_j^2/(1-z_j^2)\) and
negative beyond, so \(b_j\) rises above the limit and then falls back to
it. If \(z_j=0\), the derivative is negative. With a random cutoff, \(b_j\) and
its limit are the fixed-cutoff values averaged over the cutoff.

\paragraph{Proof of Proposition~\ref{prop:s-shape}.}
The result is Theorem 1 of \citet{FrazierPowell2010} with noise
variance \(1/\kappa\), prior variance \(1/\tau_j\), and distance
\(|\mu_j-\zeta|\). We give the short argument in our notation.
With the cutoff fixed at \(\zeta\), the slope of \(\mathcal V_j\) is the
marginal value \eqref{eq:priority-decomposition}, the product of \(b_j\)
and \(\kappa/(\tau_j+\kappa N_j)^2\). By the properties above, the
derivative of its logarithm in \(u\) is
\[
\frac{z_j^2}{2u^2}-\frac{1}{2u(1+u)}-\frac{2}{1+u}
=-\frac{4u^2+(1-z_j^2)u-z_j^2}{2u^2(1+u)} .
\]
If \(z_j>0\), the quadratic in the numerator is negative at \(u=0\), and
the product of its roots is \(-z_j^2/4<0\), so it has exactly one
positive root, \(x(z_j)\). Hence \(\mathrm{MV}_j\) increases for
\(u<x(z_j)\) and decreases for \(u>x(z_j)\), and at \(u=x(z_j)\) the
percentage rates of the two factors cancel. If \(z_j=0\), the
derivative is negative for every \(u>0\), and \(x(0)=0\). Because
\(\mathcal V_j'=\mathrm{MV}_j\), \(\mathcal V_j\) is convex where
\(\mathrm{MV}_j\) increases and concave where it decreases. Finally,
with \(y:=z^2\), \(x(z)=\big(y-1+\sqrt{y^2+14y+1}\big)/8\), which increases
in \(y\), so \(x\) increases in \(z\ge0\). Also \(x(1)=1/2\),
\(x(2)=(3+\sqrt{73})/8\approx1.44\), and
\(\sqrt{y^2+14y+1}=y+7+o(1)\) as \(y\to\infty\), so
\(x(z)/z^2\to1/4\). \hfill\(\square\)

\paragraph{Proof of Proposition~\ref{prop:envelope}.}
Fix \(\N^0\), write \(\mathcal V_j\) for \(\mathcal V_j(\cdot;\N^0)\), and
let \(\hat{\mathcal V}_j\) be its concave envelope on
\(\{0,\ldots,N_{\mathrm{tot}}\}\). Because \(\hat{\mathcal V}_j\) is
concave, its increments
\(\hat{\mathcal V}_j(n+1)-\hat{\mathcal V}_j(n)\) are nonincreasing in \(n\).
By the usual exchange argument, a sum of such functions subject to a
single budget is maximized by assigning responses one at a time to the
largest available increment, however ties are broken.
Hence
\[
\sum_j\hat{\mathcal V}_j(\hat N_j)=\max_{\N}\sum_j\hat{\mathcal V}_j(N_j)
\ \ge\ \max_{\N}\sum_j\mathcal V_j(N_j),
\]
where the inequality holds because \(\hat{\mathcal V}_j\ge\mathcal V_j\).

The envelope \(\hat{\mathcal V}_j\) is piecewise linear and equals
\(\mathcal V_j\) at the ends of each linear piece, so
\(\hat{\mathcal V}_j>\mathcal V_j\) is possible only strictly inside a
piece, where all increments are equal. Suppose the rule gives item \(j\)
the first response of one of its pieces. That increment is at least as
large as every other item's next increment, and item \(j\) comes first in
the fixed order among the items whose next increment ties with it. Within
this piece, item \(j\)'s next increment stays the same and the other
items' next increments do not change, so the rule keeps giving responses
to item \(j\) until the piece ends or the budget runs out. Hence, when the budget runs out, at most one item lies
strictly inside a linear piece, and every other item satisfies
\(\hat{\mathcal V}_j(\hat N_j)=\mathcal V_j(\hat N_j)\). Therefore
\[
\sum_j\mathcal V_j(\hat N_j)\ \ge\ \sum_j\hat{\mathcal V}_j(\hat N_j)-\max_j\eta_j
\ \ge\ \max_{\N}\sum_j\mathcal V_j(N_j)-\max_j\eta_j .
\]
Finally, each \(\mathcal V_j\) is nondecreasing, because more responses
spread \(X_j(n)\) and \((x-\zeta)^+\) is convex. The constant
\(\mathcal V_j(N_{\mathrm{tot}})\) is therefore a concave function above
\(\mathcal V_j\), so \(\hat{\mathcal V}_j\le\mathcal V_j(N_{\mathrm{tot}})\), and
\(\mathcal V_j\ge\mathcal V_j(0)\). This gives
\(\eta_j\le\mathcal V_j(N_{\mathrm{tot}})-\mathcal V_j(0)\). \hfill\(\square\)

\paragraph{The square-root rule.}
With many responses, the decision loss is close to
\(\sum_ic_i^\infty/(\tau_i+\kappa N_i)\). Minimizing this approximation
over allocations that use the whole budget gives
\begin{equation}\label{eq:sqrt-rule}
\tau_i+\kappa N_i=\max\Big(\tau_i,\ \frac{\sqrt{c_i^\infty}}{\lambda}\Big),
\end{equation}
with \(\lambda>0\) set so that the responses add up to
\(N_{\mathrm{tot}}\). We call this the \emph{square-root rule}. It needs
one number per item and no value curves, and Section~\ref{sec:numerics}
uses it as a benchmark.

\begin{corollary}[Square-root rule for large budgets]
\label{cor:sqrt-rule}
Under the conditions of Proposition~\ref{prop:sqrt-law}, the allocation
\(\N^{\mathrm{sq}}\) of the square-root rule satisfies
\(\Delta(\N^{\mathrm{sq}})/\Delta(\N^\star)\to1\) as
\(N_{\mathrm{tot}}\to\infty\).
\end{corollary}

\paragraph{Proof of Proposition~\ref{prop:sqrt-law} and Corollary~\ref{cor:sqrt-rule}.}
By Proposition~\ref{prop:loss-representation},
\(\Delta(\N)=\sum_ic_i(\N)/(\tau_i+\kappa N_i)\). Write \(\sigma_i^2=1/\tau_i\)
for the prior variance of item \(i\).

\emph{Step 1 (limits and bounds on the weights).} Along the path of
Proposition~\ref{prop:loss-representation}, \(X_t\) is normal with
covariance \(\bSigma-(1-t)A(\N)\). As \(\min_iN_i\to\infty\),
\(A(\N)\to0\), so for every \(t\) the law of \(X_t\) converges to the
prior, and the variance of \(X_{t,i}\) is at least
\(\sigma_i^2/2\) once \(a_i(\N)\le\sigma_i^2/2\). The densities
\(f_{i,t}\) are then bounded uniformly, and dominated convergence gives
\(c_i(\N)\to c_i^\infty\). For \(t\in[\tfrac12,1]\) and every allocation,
each \(X_{t,k}\) has mean \(\mu_k\) and variance between \(\sigma_k^2/2\)
and \(\sigma_k^2\). It therefore lies in
\(B:=[\min_k\mu_k-1,\,\max_k\mu_k+1]\) with probability at least
\(\Pr(|Z|\le1/\sigma_k)\) for \(Z\) standard normal, so \(\zeta_i(X_t)\in B\) with probability at
least \(p:=\prod_k\Pr(|Z|\le1/\sigma_k)>0\), and \(f_{i,t}\) is at least
some \(\delta>0\) on \(B\). Hence \(c_i(\N)\ge\underline c:=\delta p/4\)
for all \(i\) and all \(\N\).

\emph{Step 2 (a bound for every allocation).} For any weights
\(\gamma_i>0\) and any \(\N\ge0\) with \(\sum_iN_i=N_{\mathrm{tot}}\), the
Cauchy--Schwarz inequality gives
\[
\Big(\sum_i\sqrt{\gamma_i}\Big)^2
\le\Big(\sum_i\frac{\gamma_i}{\tau_i+\kappa N_i}\Big)
\Big(\sum_i(\tau_i+\kappa N_i)\Big),
\quad\text{so}\quad
\sum_i\frac{\gamma_i}{\tau_i+\kappa N_i}\ \ge\
\frac{\big(\sum_i\sqrt{\gamma_i}\big)^2}{\sum_i\tau_i+\kappa N_{\mathrm{tot}}},
\]
with equality if and only if \(\tau_i+\kappa N_i\) is proportional to
\(\sqrt{\gamma_i}\).

\emph{Step 3 (lower bound).} Fix \(\varepsilon>0\) and choose \(n_0\) such
that \(c_i(\N)\ge c_i^\infty-\varepsilon\) for all \(i\) whenever
\(\min_iN_i\ge n_0\). If some \(N_i<n_0\), then
\(\Delta(\N)\ge\underline c/(\tau_i+\kappa n_0)\), a positive constant.
Otherwise Step 2 applies with \(\gamma_i=c_i^\infty-\varepsilon\). Hence,
for \(N_{\mathrm{tot}}\) large,
\(\Delta(\N^\star)\ge\big(\sum_i\sqrt{c_i^\infty-\varepsilon}\big)^2/
(\sum_i\tau_i+\kappa N_{\mathrm{tot}})\), and letting
\(\varepsilon\to0\) gives
\(\liminf N_{\mathrm{tot}}\Delta(\N^\star)\ge(\sum_i\sqrt{c_i^\infty})^2/\kappa\).

\emph{Step 4 (upper bound, the square-root rule, and the uniform
allocation).} For \(N_{\mathrm{tot}}\) large, \(\lambda\) in
\eqref{eq:sqrt-rule} is small enough that \(\sqrt{c_i^\infty}/\lambda>\tau_i\)
for every \(i\), so the square-root rule sets
\(\tau_i+\kappa N_i\propto\sqrt{c_i^\infty}\). Every \(N_i\) then grows
without bound, so the weights converge to \(c_i^\infty\). By the
equality case of Step 2, \(N_{\mathrm{tot}}\Delta(\N^{\mathrm{sq}})\)
converges to \((\sum_i\sqrt{c_i^\infty})^2/\kappa\). This matches the
lower bound of Step 3, so \(N_{\mathrm{tot}}\Delta(\N^\star)\) has the
same limit, which also proves Corollary~\ref{cor:sqrt-rule}. For the uniform allocation, every
\(N_i\to\infty\) and \(N_{\mathrm{tot}}\Delta(\N^{\mathrm u})=N_{\mathrm{tot}}\sum_i
c_i(\N^{\mathrm u})/(\tau_i+\kappa N_{\mathrm{tot}}/L)\to
L\sum_ic_i^\infty/\kappa\).

\emph{Step 5 (shares).} By Steps 3 and 4, an optimal allocation has every
\(N_i\to\infty\) for large \(N_{\mathrm{tot}}\), because
\(\Delta(\N^\star)\le\Delta(\N^{\mathrm{sq}})\to0\) while
\(\Delta(\N)\ge\underline c/(\tau_i+\kappa n)\) whenever \(N_i<n\). Its
weights therefore converge to \(c_i^\infty\). If along some subsequence its shares stayed away from
\(\sqrt{c_i^\infty}/\sum_k\sqrt{c_k^\infty}\), the
Cauchy--Schwarz inequality in Step 2 would be strict by a margin bounded
away from zero, and its loss times \(N_{\mathrm{tot}}\) would exceed the
limit in Step 4, contradicting optimality. Dividing the two limits gives
\eqref{eq:targeting-gain}. Finally, \(L_{\mathrm{eff}}\le L\) by the
Cauchy--Schwarz inequality, and \(L_{\mathrm{eff}}\ge1\) because
\((\sum_i\sqrt{c_i^\infty})^2\ge\sum_ic_i^\infty\). \hfill\(\square\)

The decision loss thus falls like \(1/N_{\mathrm{tot}}\). For fixed
true values it would fall exponentially \citep{GlynnJuneja2004}, but
averaged over the prior it is dominated by values at which the items in
\(K\)th and \((K+1)\)th place nearly tie.

\section{Proofs for Section 5}\label{app:proofs-format}
\paragraph{Proof of Proposition~\ref{prop:eis-ranking}.}
(i) Under \eqref{eq:factorized}, the residual covariance satisfies
\(A(\N;S)=\big(\bSigma^{-1}+\diag(\kappa(S)\N)\big)^{-1}
=\big(\bSigma^{-1}+\diag(\kappa(S_0)\N')\big)^{-1}=A(\N';S_0)\) with
\(\N'=\beta(S)\N\). Hence the proxy decision loss is
the same under \((\N,S)\) and \((\N',S_0)\). The map
\(\N\mapsto\beta(S)\N\) is a bijection from the allocations with
\(\sum_iN_i\le T\,Q\,\alpha(Q,S)\) onto those with
\(\sum_iN_i'\le T\,\mathrm{EIS}(Q,S)\). Therefore
\(\Delta^\star(Q,S)=\Psi\big(T\,\mathrm{EIS}(Q,S)\big)\), where \(\Psi(b)\) is
the smallest proxy decision loss over baseline-scale allocations with
total at most \(b\). The function \(\Psi\) does not depend on the format,
and it is nonincreasing because its feasible set grows with \(b\).

(ii) The envelope rule uses the value curves \(\mathcal V_j(n)\) of
\eqref{eq:item-value}. Because the rule treats the prior as independent, item \(j\)'s
estimate \(X_j(n)\) has variance
\(\Sigma_{jj}-1/(\tau_j+\kappa(S)n)\) with \(\tau_j:=1/\Sigma_{jj}\). The law of the cutoff \(\zeta_j(X)\)
depends on the other items' variances, which depend on their responses
only through \(\kappa(S)N_i\). Written in baseline-equivalent responses
\(n'=\beta(S)n\), the curves and their concave envelopes are therefore
the same under every scale, and the slope per response under scale
\(S\) is \(\beta(S)\) times the slope per baseline-equivalent response,
which preserves the ranking by slope. Each batch of
Algorithm~\ref{alg:envelope} is the larger of \(\nu/\kappa(S)\) responses and a
fixed fraction of the planned total. The first is
\(\nu/\kappa(S_0)\) baseline-equivalent responses, so the cutoffs are refreshed at the
same baseline-equivalent totals. In the continuous relaxation, the
rule assigns real-valued responses and evaluates the curves exactly with
the same simulated draws of the cutoffs. It therefore traces the same
path of baseline-equivalent allocations under every scale and stops when
the total reaches \(T\,\mathrm{EIS}(Q,S)\). The
decision loss then depends on the format only through
\(T\,\mathrm{EIS}(Q,S)\). \hfill\(\square\)

\section{Additional Results for Section 6}\label{app:numerics}

We repeat Panel~A of Table~\ref{tab:proxy-ranking} with 1, 2, and 5
responses per item, adding the envelope-rule and square-root-rule
allocations to the ten designs. The rank correlation is 0.88 to
0.96, each value within 0.05 of the benchmark. The regret is at
most 0.2 percent,
except with one binary response per item, where it is 1.7 percent
against 0.7 percent for the benchmark. The proxy's decision loss divided
by the ordinal loss is between 0.99 and 1.04 for the envelope-rule
allocation and between 0.98 and 1.02 for the square-root-rule
allocation. Under the ordinal model, the envelope-rule allocation has
the lowest loss of the twelve designs in 17 of the 18 combinations of
configuration, scale, and budget.

In Figure~\ref{fig:num-allocation}, the standard error of a paired
difference with uniform allocation is at most 0.07 times the uniform
loss. The
ranges quoted in Section~\ref{subsec:num-allocation} and in the
introduction, such as 15 to 81 percent, are the smallest and largest
values over the three configurations and the two scales, which the
figure averages. The dashed lines are at 0.27, 0.52, and 0.78 for prior
variances 0.1, 0.3, and 1.

Table~\ref{tab:num-decomp} compares the envelope rule with the square-root,
marginal-value, minimum-share, and OCBA-m rules of Section~\ref{subsec:num-allocation}. Each entry is the
envelope rule's loss minus the other rule's, divided by the uniform
loss, and none is significantly above zero.

\begin{table}[ht]
\centering
\caption{Decision loss of the envelope rule minus that of another rule,
relative to uniform allocation. Mean (standard error) over 240
instances.}
\label{tab:num-decomp}
\small
\begin{tabular}{@{}llcccc@{}}
\toprule
Prior variance & Responses per item & Square-root & Marginal-value & Minimum-share & OCBA-m \\
\midrule
0.1 & 10 & $-$0.028 (0.002) & $-$0.011 (0.002) & $-$0.043 (0.002) & $-$0.024 (0.003) \\
 & 30 & $-$0.015 (0.002) & $-$0.030 (0.003) & $-$0.037 (0.002) & $-$0.030 (0.003) \\
 & 100 & $-$0.012 (0.002) & $-$0.115 (0.007) & $-$0.038 (0.003) & $-$0.049 (0.004) \\
 & 300 & $-$0.005 (0.003) & $-$0.292 (0.016) & $-$0.030 (0.004) & $-$0.048 (0.005) \\
\addlinespace[4pt]
0.3 & 10 & $-$0.038 (0.002) & $-$0.042 (0.003) & $-$0.046 (0.002) & $-$0.053 (0.003) \\
 & 30 & $-$0.023 (0.003) & $-$0.111 (0.005) & $-$0.033 (0.002) & $-$0.062 (0.003) \\
 & 100 & $-$0.012 (0.004) & $-$0.302 (0.010) & $-$0.029 (0.004) & $-$0.059 (0.005) \\
 & 300 & $-$0.006 (0.004) & $-$0.619 (0.024) & $-$0.019 (0.004) & $-$0.037 (0.006) \\
\addlinespace[4pt]
1 & 10 & $-$0.041 (0.003) & $-$0.021 (0.003) & $-$0.033 (0.002) & $-$0.069 (0.003) \\
 & 30 & $-$0.007 (0.003) & $-$0.078 (0.005) & $-$0.012 (0.002) & $-$0.036 (0.003) \\
 & 100 & $-$0.004 (0.003) & $-$0.151 (0.012) & $-$0.007 (0.002) & $-$0.032 (0.005) \\
 & 300 & 0.001 (0.003) & $-$0.212 (0.028) & $-$0.005 (0.003) & $-$0.029 (0.006) \\
\bottomrule
\end{tabular}
\end{table}

Table~\ref{tab:num-wrongprior} gives the results of the wrong-prior
experiment in Section~\ref{subsec:num-robust}. The shuffled items and
the three better items are drawn anew in each replication. Each entry
averages the random and crowded configurations and the two scales, with
30 instances and 100 replications each, and is relative to uniform
allocation with the prior as given. The standard error of a paired
difference with uniform allocation is at most 0.07 times the uniform
loss, except in the rows with shuffled prior means, where it reaches
0.29. With two waves, the
envelope rule's loss relative to uniform allocation under the four wrong
priors is 0.46 to 0.78 with prior variance \(1\) and 0.50 to 2.02 with
\(0.3\) (random configuration, binary ratings, 30 and 100
responses per item, 20 instances with 10 replications each).

\begin{table}[ht]
\centering
\caption{Ordinal decision loss relative to uniform allocation under
right and wrong priors.}
\label{tab:num-wrongprior}
\small
\begin{tabular}{@{}l*{8}{S[table-format=1.2]}@{}}
\toprule
 & \multicolumn{4}{c}{Prior variance 0.3} & \multicolumn{4}{c}{Prior variance 1} \\
\cmidrule(lr){2-5}\cmidrule(l){6-9}
Responses per item & {10} & {30} & {100} & {300} & {10} & {30} & {100} & {300} \\
\midrule
\multicolumn{9}{@{}l}{\emph{Envelope rule, prior as given}} \\
Right prior                          & 0.46 & 0.47 & 0.48 & 0.49 & 0.65 & 0.70 & 0.74 & 0.73 \\
True variance twice the prior's      & 0.62 & 0.72 & 0.84 & 0.90 & 0.78 & 0.82 & 0.87 & 0.91 \\
True variance four times the prior's & 0.92 & 1.29 & 1.86 & 2.42 & 0.89 & 0.97 & 1.00 & 1.11 \\
Prior means shuffled                 & 1.22 & 2.13 & 4.42 & 7.74 & 0.97 & 1.15 & 1.53 & 1.75 \\
Three items better                   & 0.79 & 1.06 & 1.43 & 1.71 & 0.95 & 1.00 & 1.22 & 1.32 \\
\addlinespace[3pt]
\multicolumn{9}{@{}l}{\emph{Envelope rule, prior variance doubled}} \\
Right prior                          & 0.54 & 0.55 & 0.55 & 0.55 & 0.71 & 0.73 & 0.79 & 0.77 \\
True variance twice the prior's      & 0.52 & 0.57 & 0.62 & 0.65 & 0.72 & 0.74 & 0.81 & 0.81 \\
True variance four times the prior's & 0.59 & 0.68 & 0.76 & 0.86 & 0.77 & 0.80 & 0.81 & 0.83 \\
Prior means shuffled                 & 0.86 & 1.18 & 1.70 & 2.63 & 0.83 & 0.95 & 1.13 & 1.08 \\
Three items better                   & 0.59 & 0.64 & 0.66 & 0.68 & 0.73 & 0.74 & 0.81 & 0.80 \\
\bottomrule
\end{tabular}
\end{table}

\section{Retention Associations in an Industrial Curation Log}\label{app:retention-drivers}

This appendix summarizes an observational retention study of an
industrial subscription curation service, conducted on the service's
order logs by one of the authors, a former employee, as part of a
research collaboration with the service. It supports the statement in
Section~\ref{sec:intro} that repeating items is associated with lower
retention. The unit of analysis is an order, that is, a shipped box.
Retention is whether the customer orders again. We measure it with three labels, a reorder within
the customer's usual interval, a reorder within thirty days, and
no cancellation shortly after the order. The
estimation sample covers about \(52{,}000\) orders placed between July
and November 2025. We fit a separate logistic regression for each label, with
standardized continuous features, so each estimate is an odds ratio per one
standard deviation.

The cross-order repeat rate is the share of the
box's items that also appeared in the customer's previous four orders,
computed overall and separately for recipe and occasional items.
Premium intensity is the average credit points per item. Within-box
duplication density is the number of duplicated items divided by box
size, and the duplication spike is the change in duplication relative
to the customer's trailing mean. Tenure enters as the logarithm of one
plus the customer's lifetime order count.

The odds ratio for the
repeat rate varies across customer segments defined by total credit and box
size. The association is stronger with more credit and slightly weaker
in larger boxes. Only aggregate, relative quantities are reported, with the
service's permission, and the underlying log cannot be released. The regressions condition on customer tenure,
but repeat rates and box economics can still correlate with unmeasured
engagement, so we use these estimates only as motivation.

\begin{table}[ht]
\centering
\caption{Retention associations in an industrial subscription curation
log. Odds ratios per one standard deviation from order-level logistic
regressions. Conclusions are the same under the three retention labels.}
\label{tab:retention-drivers}
\small
\begin{tabular}{@{}lcl@{}}
\toprule
Box-composition feature (per \(+1\) SD) & Odds ratio & Note \\
\midrule
Cross-order repeat rate & 0.63--0.88 & range across customer segments \\
Repeat of recipe items & \(\approx 0.80\) & significant \\
Repeat of occasional items & \(\approx 0.92\) & not significant \\
Order size & 1.23 & \\
Premium intensity (points per item) & 0.97 & \\
Within-box duplication density & 0.98 & \\
Duplication spike vs.\ customer's trailing mean & 0.97 & \\
\bottomrule
\end{tabular}
\end{table}

In Table~\ref{tab:retention-drivers}, repeated items are associated with
lower retention, most strongly for repeated recipes. Larger boxes are associated with
higher retention, and higher premium intensity with lower retention.
Repetition and box economics concern which items are candidates, not
the item values of
Section~\ref{sec:model}, so the evidence supports only the narrower
premise that what a box contains is associated with retention.

\end{document}